\documentclass[aps,amsmath,amsfonts,11pt]{revtex4}
\usepackage{graphicx}
\usepackage{epstopdf}
\usepackage{epsfig,graphicx}
\usepackage[english]{babel}
\usepackage{amsfonts}
\usepackage{amsmath}
\usepackage{latexsym}
\usepackage{graphics,bm}
\usepackage{dcolumn}
\usepackage{bm}
\usepackage{rotating}

\begin{document}

\title{Nonlinear optical response properties of a quantum dot embedded in a semiconductor microcavity : possible applications in quantum communication platforms
}

\author{ Vijay Bhatt$^{1}$, Sabur A. Barbhuiya$^{2}$, Pradip K. Jha$^{1}$, Aranya B. Bhattacherjee$^{2}$}

\address{$^{1}$Department of Physics, DDU College, University of Delhi, New Delhi 110078, India }
\address{$^{2}$Department of Physics, Birla Institute of Technology and Science, Pilani, Hyderabad Campus, Hyderabad - 500078, India}

\begin{abstract}
We theoretically investigate optical bistability, mechanically induced absorption (MIA) and Fano resonance of a hybrid system comprising of a single quantum dot (QD) embedded in a solid state microcavity interacting with the quantized cavity mode and the deformation potential associated with the lattice vibration. We find that the bistability can be tuned by the QD-cavity mode coupling. We further show that the normalized power transmission displays anomalous dispersion indicating that the system can be used to generate slow light. We also demonstrate the possibility of using the system as all optomechanical Kerr switch. 
\end{abstract}

\maketitle

\section{Introduction}
Quantum optomechanics in the recent past has rapidly emerged as a major research area \citep{aspes1,kippen}. The photon and phonon interactions via radiation pressure is exploited in cavity optomechanics, which usually comprises of a mechanical oscillator and an optical cavity\citep{kippen,mar,aspes2}.

The optomechanic research is primarily conducted at meso/nanoscales \citep{aspes3} since at these scales, the strong optomechanical coupling can generate numerous interesting phenomena, viz., optical bistability \citep{arcizet,letter,jiang,park}, optomechanically induced transparency (OMIT) \citep{weis,safavi,karuza,kronwald,ma,hou,xiong}, optomechanically induced absorption (OMIA) \citep{hou,xiong,hocke}, normal-mode splitting \citep{dobrindt,groblacher,kamran,huang1,bhatt1,he,huang2}, and side band cooling \citep{teufel,marquardt,schliesser,riviere,chan,Pierre}.

Introducing additional components into the optomechanical system can provide new handle to control the optomechanical effects \citep{yella,jiao,lin,ott,miro,fano,verellen}. Recently solid state based analog of optomechanics has also been developing rapidly \citep{naumann,mahajan,stufler,dezf}. Such solid state based optomechanical systems can be realized using distributed Bragg reflectors (DBRs). Semiconductor quantum dots(QDs) have emerged as a new class of hybrid quantum devices which are confined within the optical micro-cavities owing to their large density of states and high tunability \citep{vahala1,ali,vahala2,khitrova,tang,majumdar,bhatt2,majumdar2,bhatt3}.

Because of its high-quality factor and small and finite volume, photonic crystal micro-cavities are preferred. One of the advantages of solid state based systems is easy fabrication and integration into large scale array for possible quantum network plateforms.

The basic idea of solid state analog of optomechanics, was extended to a single quantum dot which was shown to exibit the feature of the optomechanical system due to it's small size \citep{li}. In this system, the QD's excitonic resonance plays the role of optical cavity, while the lattice vibrations are present in place of mechanical oscillator. The deformation potential serves to couple the exciton and phonons.

In the present article, fabricated single quantum dot in a solid state micro-cavity interacting with the quantized light mode and the deformation potential associated with the lattice vibrations was investigated.

We analyze in detail the occurrence of optical bistability, mechanically induced absorption (MIA) and Fano resonance and their dependence on exciton-pump detuning, exciton-lattice vibration coupling, and exciton-cavity mode coupling. We find that the appearance of the bistability can be tuned effectively by the coupling strength of QD-cavity mode. The coupling between the QD-lattice vibrations is distinctly visible in the optical response properties of the system. We also show the appearance of anomalous dispersion in the transmitted signal due to exciton-phonon coupling. These results strongly indicate the possibility of using this hybrid system in quantum information processing plateforms.    

 \section{THEORETICAL MODEL}
 The model proposed in this article is shown in Fig.1. Here we consider a single semiconductor quantum dot (QD) inside a semiconductor micro cavity. The micro-cavity can be fabricated by a set of distributed Bragg reflectors (DBR). Light confinement along the longitudinal and transverse direction in the DBR can be achieved by known techniques \citep{gudat}. DBR mirrors are generally quarter wavelength thick, a high and lower refractive index which leads to tunable reflectivity \citep{choy}. 
 
  \begin{figure}[h]
 	\includegraphics [scale=0.8]{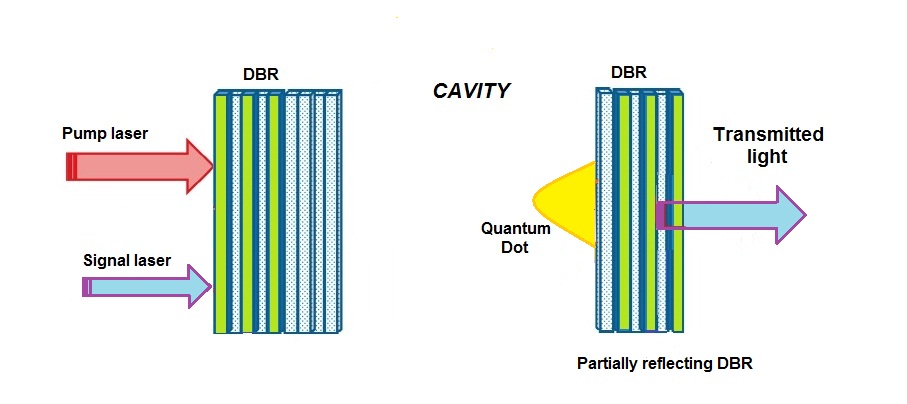}\\
 	\caption{ Schematic display of the configuration discussed in the text. DBR mirrors with the cavity is shown in figure. The quantum dot interacts with the cavity. White and Green colored strips represent the GaAs and AlGaAs layers respectively.}
 	\end{figure}   
    
 The optomechanics is introduced into the system through the coupling of cavity mode with the lattice vibrations \citep{li}. The quantum dot (QD) is considered as a system of two-level with $\lvert 0\rangle ( \lvert 1\rangle$) depicting the ground (excited(single exciton)) state. The QD can be characterized by the pseudospin operators $\sigma_\pm$ and $\sigma_{z}$.
 
 The Hamiltonian describing the QD is written as  $\hat{H}_{ex}= \hbar\omega_{ex}\sigma_{z}$, with exciton frequency ($\omega_{ex}$). The Hamiltonian for the cavity mode is described by $\hat{H}_{cav}= \hbar\omega_{c}a^{\dagger}a$, where $\omega_{c}$ is the frequency of the cavity mode and $a(a^{\dagger})$ is the cavity photon annihilation(creation) operator. The interaction between the cavity mode and the QD in the dipole approximation is denoted by $\hat{H}_{ex-cav}= \hbar g (\sigma_{+}a + a^{\dagger}\sigma_{-})$, where $g$ is QD-cavity coupling strength. The Hamiltonian $\hat{H}_{ph}= \sum_{\vec{k}}\hbar\omega_{\vec{k}} b_{\vec{k}}^{\dagger} b_{\vec{k}} $ , with vibrational phonon frequency ($\omega_{\vec{k}}$) and $b_{\vec{k}}^{\dagger} (b_{\vec{k}})$ is the phonon creation (annihilation) operator corresponding to phonon momentum $\hbar\vec{k}$ .

 The total Hamiltonian for this coupling system can be provided in the presence of a strong (pump) field and a weak(signal) field as,
 
  \begin{equation}
 \hat{H}=\hat{H}_{ex} + \hat{H}_{cav} + \hat{H}_{ex-cav} + \hat{H}_{ph} + \hat{H}_{ex-ph} + \hat{H}_{p}+ \hat{H}_{s},
 \end{equation}                     
 
  \begin{equation}
 \hat{H}_{ex-ph}=\hbar \sigma_{z}\sum_{\vec{k}}M_{\vec{k}}(b^{\dagger}_{\vec{k}} + b_{\vec{k}})
 \end{equation}
 
  \begin{equation}
 \hat{H}_{p}+\hat{H}_{s}=-i\hbar E_{p}(a e^{i \omega_{p}t} - a^{\dagger} e^{-i \omega_{p}t})-i\hbar E_{s}(a e^{i \omega_{s}t} - a^{\dagger} e^{-i \omega_{s}t})
 \end{equation}
 
Here $\hat{H}_{ex-ph}$ shows the coupling of phonon-exciton and $M_{\vec{k}}$ denoting it's coupling strength. $\hat{H}_{p}(\hat{H}_{s})$ describes the strong pump (weak signal) $E_{p}$ ($E_{s}$) as the slowly varying envelop of the strong pump (weak signal).

In the rotating frame with frequency of pump field ($\omega_{p}$), the total Hamiltonian is written as,

 \begin{eqnarray}
\hat{H}_{rot}&=&\hbar \Delta_p \sigma_{z} + \hbar \delta_c a^{\dagger}a + \hbar g(\sigma_{+}a +a^{\dagger}\sigma_{-}) -i \hbar E_{p}(a - a^{\dagger}) -i \hbar E_{s}(a e^{i\delta t} - a^{\dagger} e^{-i\delta t}) \nonumber \\
&+& \sum_{\vec{k}}{\omega_{\vec{k}}(b_{\vec{k}}^{\dagger}b_{\vec{k}})} + \hbar \sigma_{z}\sum_{\vec{k}}M_{\vec{k}}(b_{\vec{k}}^{\dagger} + b_{\vec{k}}),
\end{eqnarray}

 where $\Delta_p = \omega_{ex} - \omega_{p}$ denotes pump-exciton detuning, $\Delta_c = \omega_{c} - \omega_{p}$ denotes cavity-pump detuning and $\delta= \omega_{p}-\omega_{s}$ is denoting the signal-pump detuning.

 It has been shown earlier that a single QD is equivalent to a cavity optomechanical system \citep{majumdar2,ali}. Now the corresponding Heisenberg-Langevin equation are obtained in the following manner.

  \begin{equation}  
 \frac{d\sigma_{z}}{dt}=-\Gamma_{1}\left( \sigma_{z} + 1 \right)  - i g\left( a \sigma_{+} - a^{\dagger}\sigma_{-}\right),
 \end{equation}

 \begin{equation}
 \frac{d\sigma_{-}}{dt}=-(\Gamma_{2} + i \Delta_p)\sigma_{-} - iq\sigma_{-} + 2iga\sigma_{z} + F_{n},
 \end{equation}

 \begin{equation}
 \frac{da}{dt}=-(i\Delta_c + \kappa_{c})a - ig\sigma_{-} + E_{p} + E_{s}e^{-i\delta t},
 \end{equation}

 \begin{equation}
 \frac{d^{2}q}{dt^{2}} + \gamma_{q}\frac{dq}{dt} + 2\eta\omega^{3}_{\vec{k}} \sigma_{z} - \omega^{2}_{\vec{k}} q=0,
 \end{equation}

 where $\eta$=$\sum_{\vec{k}}\frac{M^{2}_{\vec{k}}}{\omega^{2}_{\vec{k}}}$ is the Huang-Rhys factor which represents the coupling of exciton-phonon \citep{huang}, $\Gamma_{1}$ is exciton spontaneous emission rate and $\Gamma_{2}$ is dephasing rate. q=$\sum_{\vec{k}}M_{\vec{k}}\left( b_{\vec{k}} + b^{\dagger}_{\vec{k}}\right)$ is phonon's position operator. $F_{n}$ is noise operator, with zero mean $<F_{n}>=0$, $\gamma_{q}$ is the phonon decay rate and $\kappa_{c}$ is photon decay rate in the cavity.

 Here we are looking the mean response of the coupled system to the signal field and hence we do not consider quantum fluctuations. Using   $<q \sigma_{-}>$ = $<q> <\sigma_{-}>$, we ignore entanglement between the phonon and exciton degree of freedom. The above equations are then rewritten as,

  \begin{equation}  
 \frac{d<\sigma_{z}>}{dt}=-\Gamma_{1}\left( <\sigma_{z}> + 1 \right)  - i g\left(< a ><\sigma_{+}> - <a^{\dagger}><\sigma_{-}>\right) 
 \end{equation}

 \begin{equation}
 \frac{d<\sigma_{-}>}{dt}=-(\Gamma_{2} + i \Delta_p)<\sigma_{-}> - i<q><\sigma_{-}> + 2ig<a><\sigma_{z}> + <F_{n}>
 \end{equation}

 \begin{equation}
 \frac{d<a>}{dt}=-\left( i\Delta_c + \kappa_{c}\right) <a> - ig<\sigma_{-}> +E_{p} + E_{s}e^{-i\delta t}
 \end{equation}

 \begin{equation}
 \frac{d^{2}<q>}{dt^{2}} + \gamma_{q}\frac{d<q>}{dt} + 2\eta\omega^{3}_{\vec{k}}<\sigma_{z}> - \omega^{2}_{\vec{k}}<q>=0
 \end{equation}

 In order to solve Equation. (9)-(12), we make the ansatz \citep{xu,boyd}
 
 $<\sigma_{-}(t)>$= $\sigma_{0} + \sigma_{+}e^{-i\delta t} + \sigma_{-}e^{i\delta t}$

 $<\sigma_{z}(t)>$= $\sigma^{z}_{0} + \sigma^{z}_{+}e^{-i\delta t} + \sigma^{z}_{-}e^{i\delta t}$

 $<a(t)>= a_{0} + a_{+}e^{-i\delta t} + a_{-}e^{i \delta t}$

 $<q(t)>= q_{0} + q_{+}e^{-i\delta t} + q_{-}e^{i \delta t}$

  By substituting these equation into equation (9)-(12) and to work with the lowest order of $E_{p}$, but to all orders in $E_{s}$, we deduce the linear optical susceptibility as,

  \begin{equation}
  \chi^{(1)}_{eff}=\frac{ \sigma_{+}}{E_{s}}=\frac{\chi(\omega_{\vec{k}})}{\Gamma_{2}}, 
  \end{equation}

  where the susceptibility(dimensionless) is given by:-

 \begin{equation}
 \chi(\omega_{\vec{k}})=\left(2\omega_{\vec{k}0}\eta \zeta_{1}(\omega_{\vec{k}})C_{1} + 2g_{0}D_{1}\right)  \left(\frac{-(2g^{3}_{0}C_{2}w_{0} - i g_{0}C_{2}A_{1}M_{1} + 2 i g^{2}_{0}D_{2}A_{1} w_{0})}{\Phi_{1} A^{2}_{1} M^{2}_{1}}\right) +\frac{2g_{0}w_{0}}{A_{1}M_{1}}.
 \end{equation}

 In a similar manner, we deduce the nonlinear optical susceptibility as:-

\begin{equation}
\chi^{(3)}_{eff}=\frac{ \sigma_{-}}{3 E_{s}^{*} E^{2}_{p}}=\frac{\chi^{3}(\omega_{\vec{k}})}{\Gamma^{3}_{2}},
\end{equation}

and the nonlinear optical susceptibility(dimensionless) as:-

\begin{equation}
\chi^{3}(\omega_{\vec{k}})=\left(2\omega_{\vec{k}0}\eta \zeta_{2}(\omega_{\vec{k}})C_{1} + 2g_{0}D_{1}\right)  \left(\frac{-(2g^{3}_{0}C_{1}w_{0} - i g_{0}C_{1}A_{2}M_{2} + 2 i g^{2}_{0}D_{1}A_{2} w_{0})}{\Phi_{2} A^{2}_{2} M^{2}_{2}N_{2}}\right).
\end{equation}

 We can determine the population inversion($w_{0}$) of the exciton by the equation as follows:-

 \begin{equation}
 -(2 w_{0} +2) - i g_{0}(D_{1}C_{2} - D_{2}C_{1})=0,
 \end{equation}

 From the standard input-output theory \citep{walls}, the transmitted output field through the coupled system can be obtained.[See Appendix-C]

 \begin{equation}
 T=\left|1-\frac{\sqrt{2 \kappa} a_{+}}{E_{s}}\right|
 \end{equation}
 
 and from Eq. (C1), we can get
 
 \begin{equation}
 a_{out+} =\sqrt{2 \kappa}a_{+}
 \end{equation}
 
 	where dimensionless parameters w.r.t $\Gamma_{2}$ are defined as:-
 
 $\gamma_{q0}=\frac{\gamma_{q}}{\Gamma_{2}}, \delta_{0}=\frac{\delta}{\Gamma_{2}},  g_{0}=\frac{g}{\Gamma_{2}},  \omega_{\vec{k}0}=\frac{\omega_{\vec{k}}}{\Gamma_{2}}, w_{0}=\sigma^{z}_{0},  \tau_{1}=2\Gamma_{2},  w_{0}=\frac{w}{\Gamma_{2}},  \Delta_{c0}=\frac{\Delta_c}{\Gamma_{2}},  \Delta_{p0}=\frac{\Delta_{p}}{\Gamma_{2}}$,

 $ \kappa_{c0}=\frac{\kappa_{c}}{\Gamma_{2}}, E_{p0}=\frac{E_p}{\Gamma_{2}}$ .

 	\section{Controllable optical bistability}   
 
 In this regime, we will illustrate how the bistability appearing in the population inversion $w_{0}$ can be controlled by modulating different system parameters. It is clearly obvious from equation (17) that the steady state value $w_{0}$ for the population inversion is a third-order polynomial equation, thus there are most three real roots.

 Radiation pressure induces dynamic backaction due to which optical bistability arises within the cavity of finite decay time \citep{vengala,brenn,purdy,babu,ghobadi,sete,dalafi,jiang,groblacher}. We will discuss here that how the strength of QD and cavity coupling ($g_{0}$) and pump strength influences the bistability.

 Figure 2(a) shows the graph of the population inversion $w_{0}$ verses pump-exciton detuning $\Delta_{p0}$ having steady-state values for different values of pump strength $E_{p0}$, keeping $g_{0}$ fixed at 0.1. All frequencies are dimensionless w.r.t dephasing rate.
 
 \begin{figure}[htb]
 	\centering
 	\begin{tabular}{@{}cccc@{}}
 	\includegraphics[width=.50\textwidth]{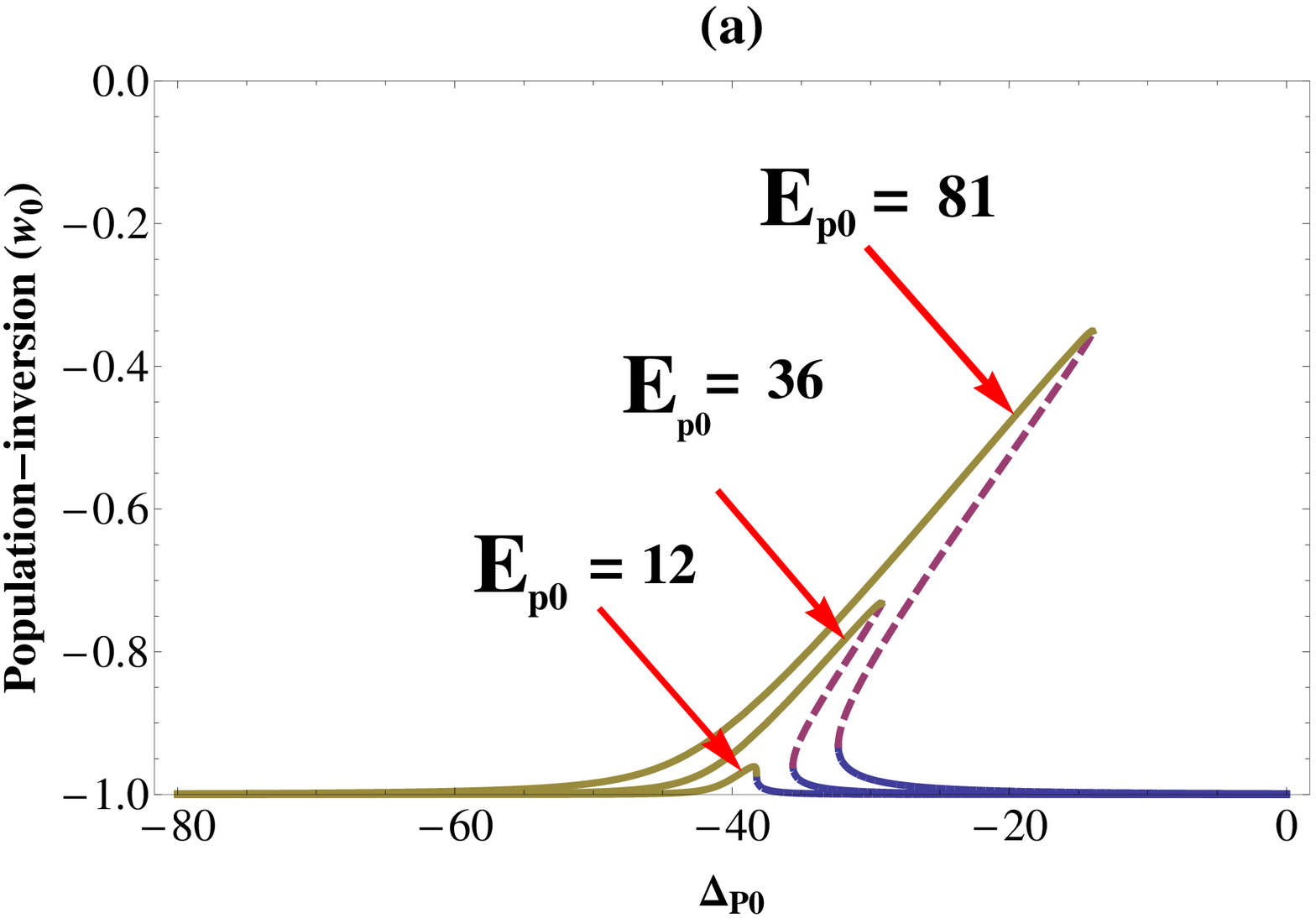} &
	\includegraphics[width=.50\textwidth]{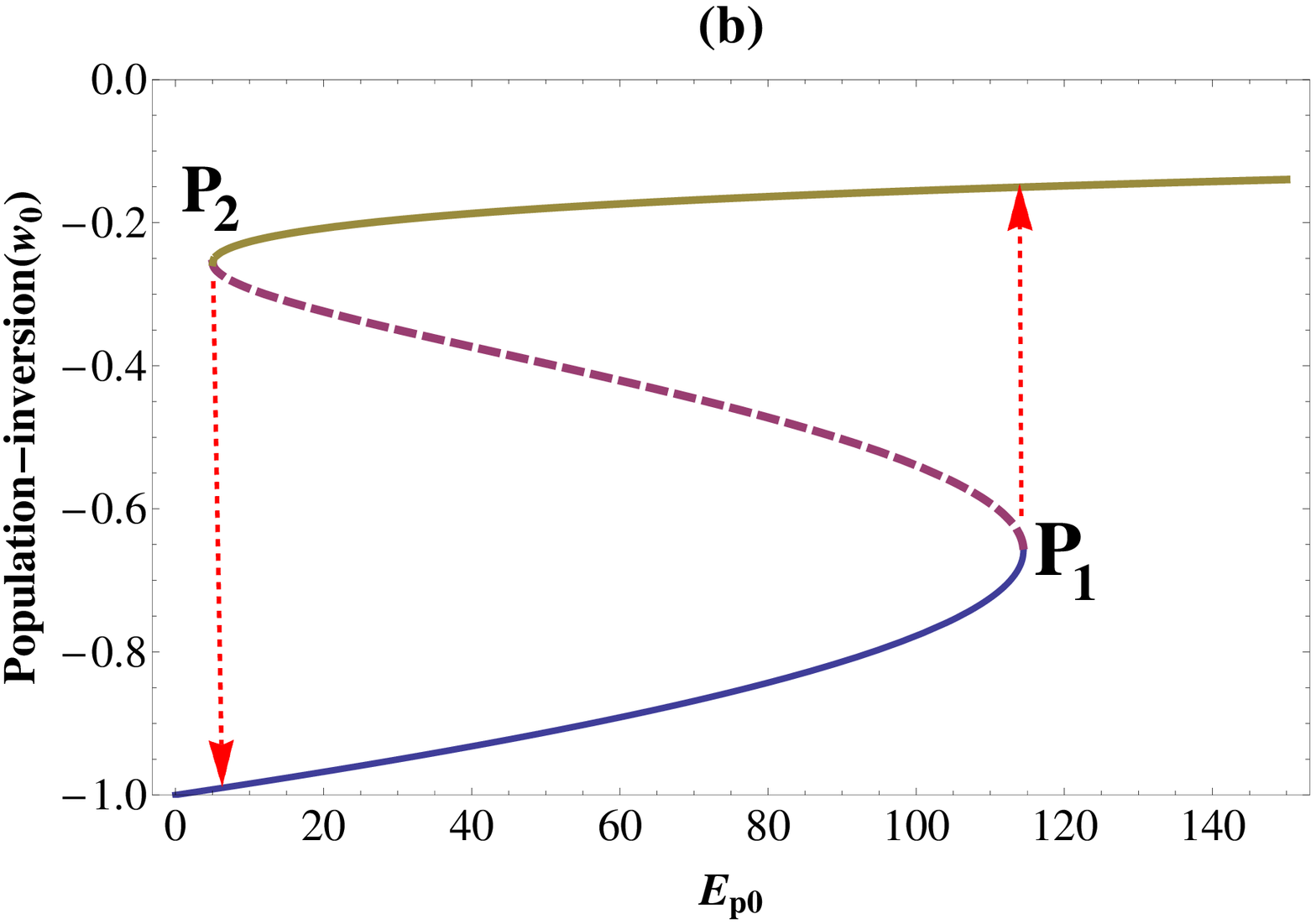} \\
 	\end{tabular}
 
 \caption{(a)- Population inversion $w_{0}$ w.r.t pump exciton detuning for different values of pump field. Parameters used for graph are-$E_{p0}$=(81,36,12), $\eta$=0.2, $g_{0}=0.1$, $\omega_{\vec{k}0}$=100, $\Delta_{c0}$=10, $\kappa_{c0}$=1.35 (b)- The graph of Population inversion $w_{0}$ verses pump field. Parameters used for graph are-$\Delta_{c0}$=0.8, $\Delta_{p0}$=-8, $g_{0}=1$, $\eta$=0.2, $\omega_{\vec{k}0}$=100, $\kappa_{c0}$=1.35. (All these parameters are dimensionless w.r.t $\Gamma_{2}$)}
\end{figure}
 
   Increasing the pump strength $E_{p0}$ results in a sharper bistable behavior. We also note that for larger pump intensity, bistability occurs at higher pump-exciton detuning.

 We plot the stationary value of $w_{0}$ w.r.t the power of the pump field $E_{p0}$ as shown in Fig. 2(b). The S-shaped behavior exhibits the bistable behavior when the largest and smallest roots of $w_{0}$ are stable, and the middle one is unstable.
 
 As the pump power is increased gradually, the population inversion moves along the lower stable branch. When this curve is reaching the first bistable point $P_{1}$, it jumps to the upper stable branch and continues on that branch. In a similar manner, decreasing the pump power, there is a gradual decrease in $w_{0}$ along the upper branch. Upon reaching the second point of  bistable $P_{2}$, it jump down to the lower stable branch and continues to decrease along that with decrease in pump power further.
 
  \begin{figure}[htb]
 	\centering
 	\begin{tabular}{@{}cccc@{}}
 	\includegraphics[width=.50\textwidth]{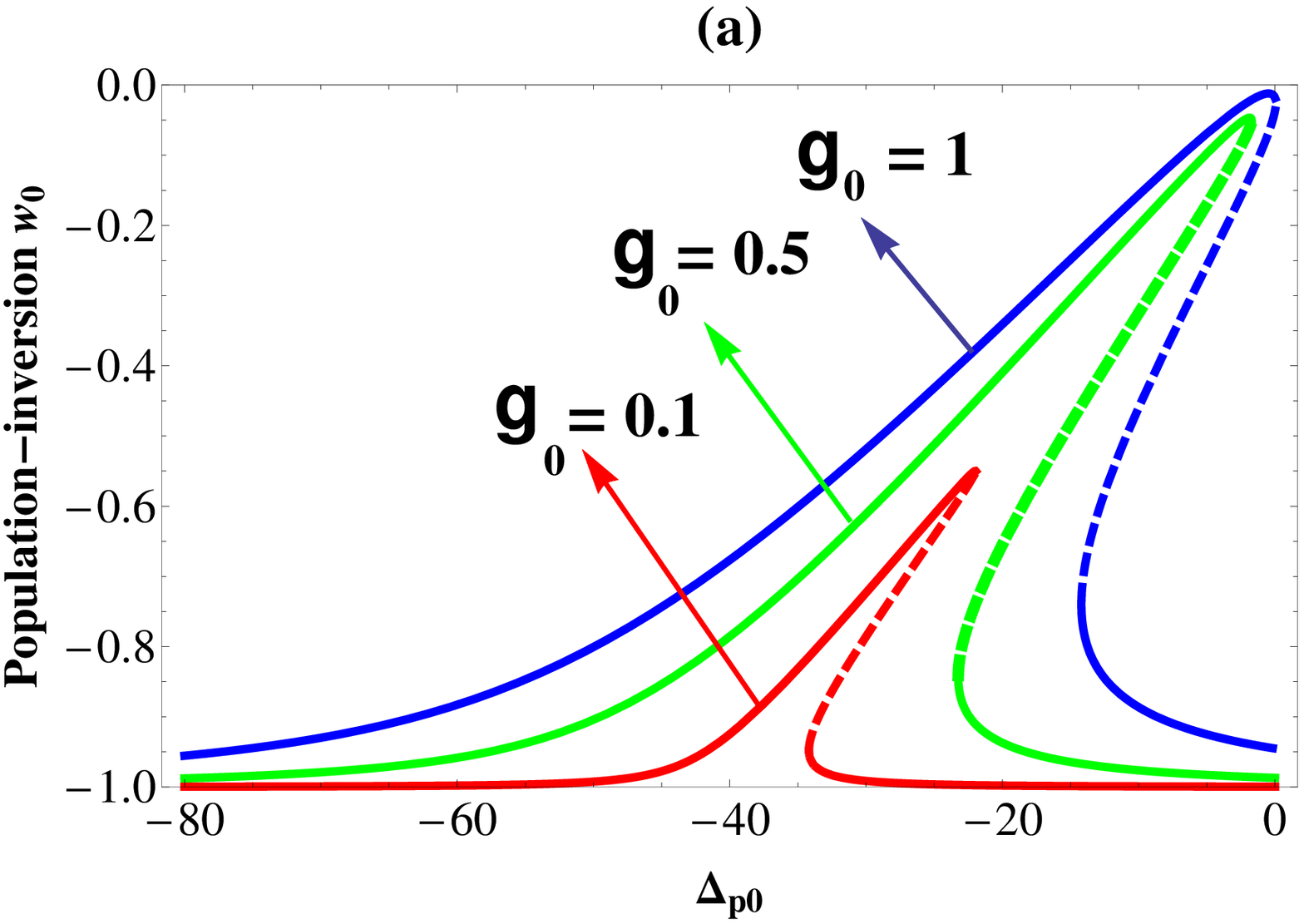} &
 	\includegraphics[width=.50\textwidth]{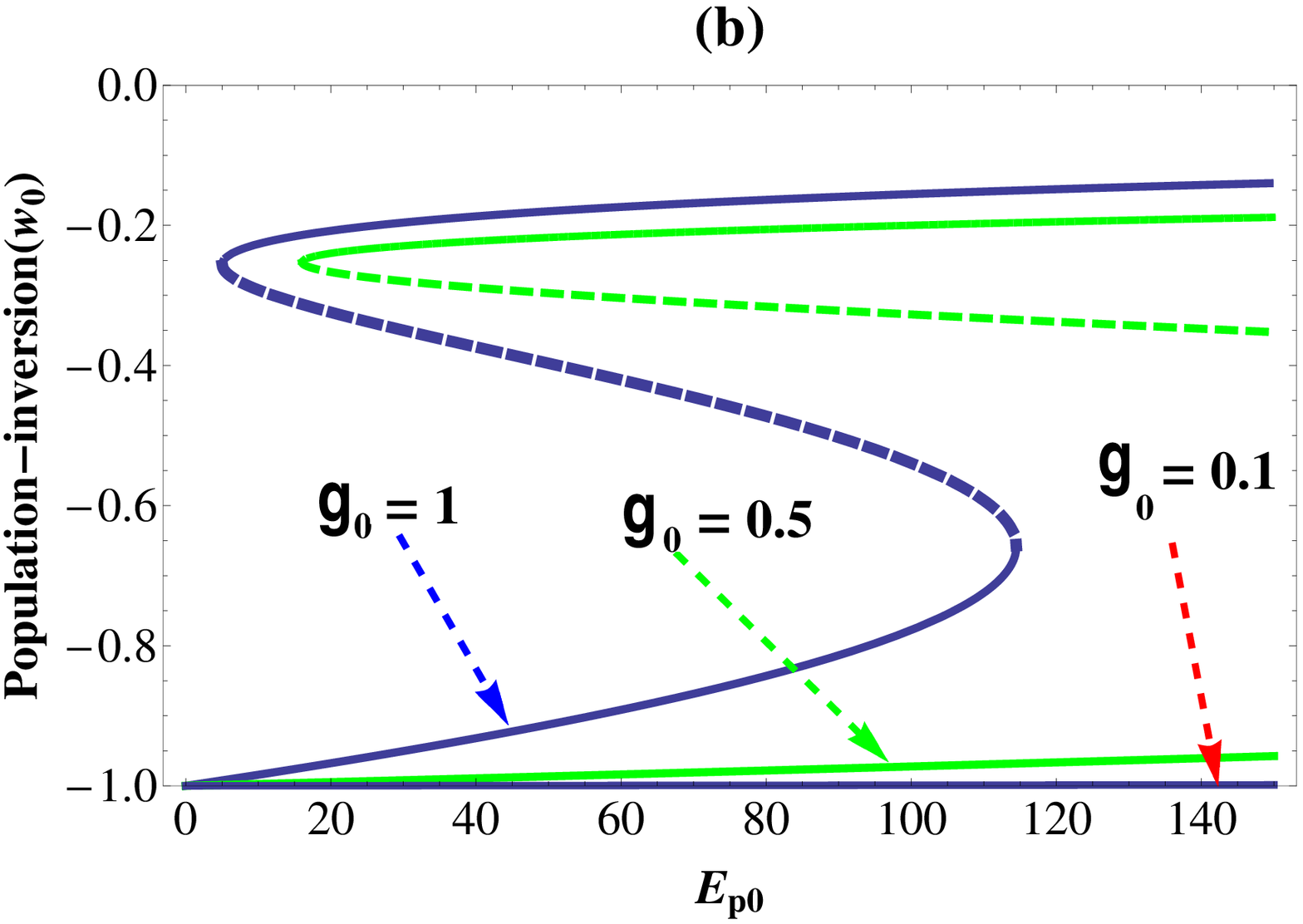}\\
 	\end{tabular}
 	\caption{(a)-Population inversion $w_{0}$ w.r.t pump-exciton detuning ($\Delta_{p0}$); Here $E_{p0}$=64, $\eta$=0.2, $\omega_{\vec{k}0}$=100, $\Delta_{c0}$=10, $\kappa_{c0}$=1.35 (b)-The graph of Population inversion $w_{0}$ verses pump field. Parameters are used for graph are-$\Delta_{c0}$=0.8, $\Delta_{p0}$=-8, $\eta$=0.2, $\omega_{\vec{k}0}$=100, $\kappa_{c0}$=1.35. (All these parameters are dimensionless w.r.t $\Gamma_{2}$)} 
 \end{figure}

 In Fig. 3(a), we plot the graph of $w_{0}$ verses pump-exciton detuning for three different values of $g_{0}$. Increasing the QD-cavity mode coupling results in the bistable behavior becoming more prominent. For low values of $g_{0}$ and $E_{p0}$, the bistability disappears. Fig. 3(b) displays the typical S-shaped bistable curves for various $g_{0}$ values. We note that, with decreasing $g_{0}$, bistability occurs at a higher value of pump strength $E_{p0}$ and for very low value of $g_{0}$, bistability disappears.

 From the above discussion, we can conclude that a fine control over the coupling of QD-cavity mode($g_{0}$) and pump strength $E_{p0}$ is required to tune the bistability which can be used for designing all-optical switches, logic-gate devices and memory devices.

 \section{Tunable MIA and Fano resonance}
 In this section, we demonstrate mechanically induced absorption (MIA) in our coupled QD-phonon system. In MIA the mechanics is introduced due to lattice vibrations while in OMIT/OMIA radiation pressure is responsible for the mechanics. It is well known that OMIT/OMIA has a sharp symmetrical profile. On the other hand, Fano-resonance has a sharp asymmetrical spectral profile \citep{fano,lin,ott,miro}. Fano-resonance's sharp asymmetrical line shape can be used to design sensitive optical switch since any variation in the parameters of the system can lead to subsequent change in the phase and amplitude of the transmitted light. We now explore the possibility to tune the MIA and Fano-resonance using the system parameters.
 
 Figure 4.(a) and 4.(b) shows the signal absorption spectrum w.r.t $\delta_{0}$ without ($ \eta$=0) and with ($\eta$=0.02) the coupling between exciton and phonon.
 
 \begin{figure}[htb]
 	\centering
 	\begin{tabular}{@{}cccc@{}}
 		\includegraphics[width=.50\textwidth]{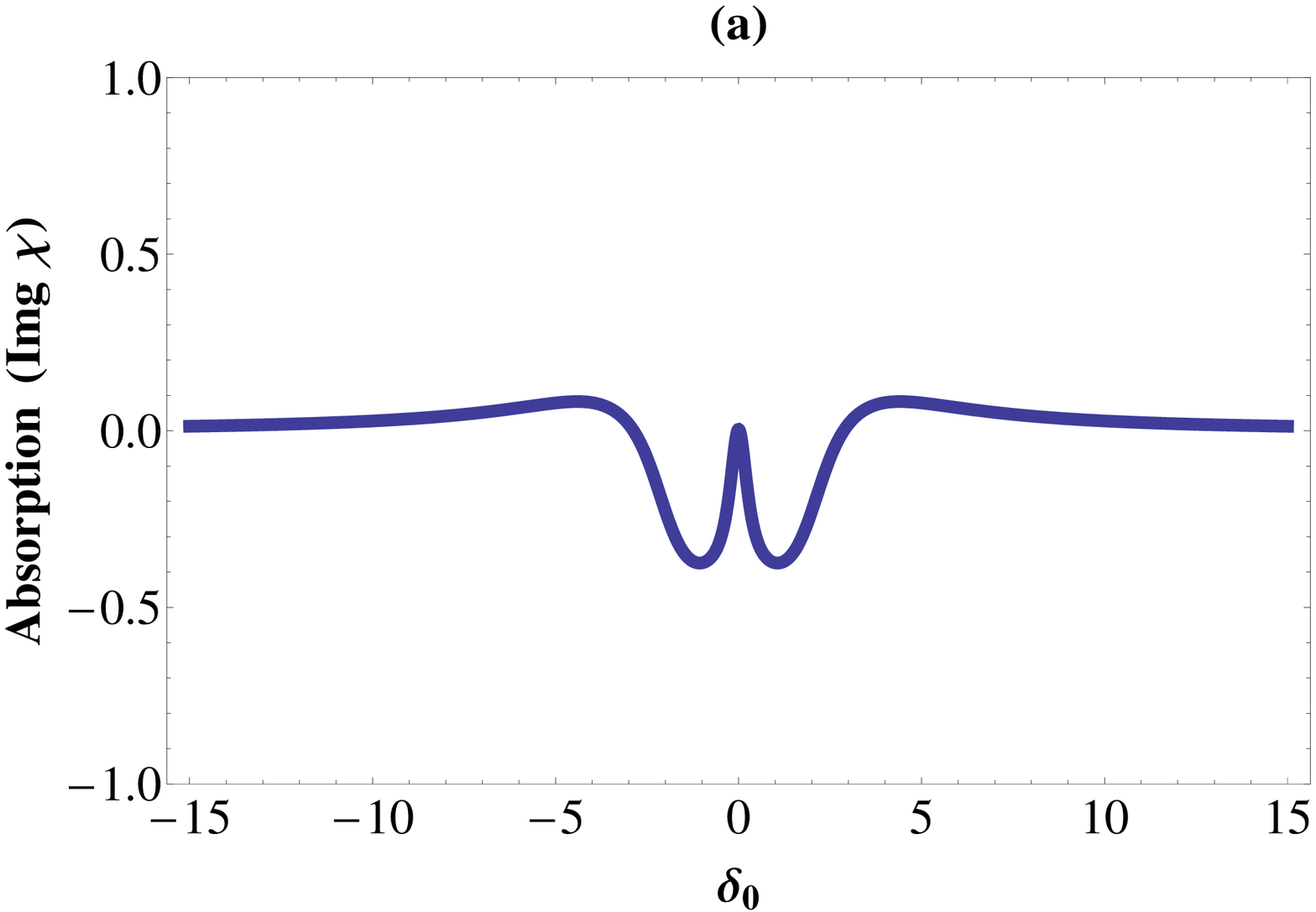}&
 		\includegraphics[width=.50\textwidth]{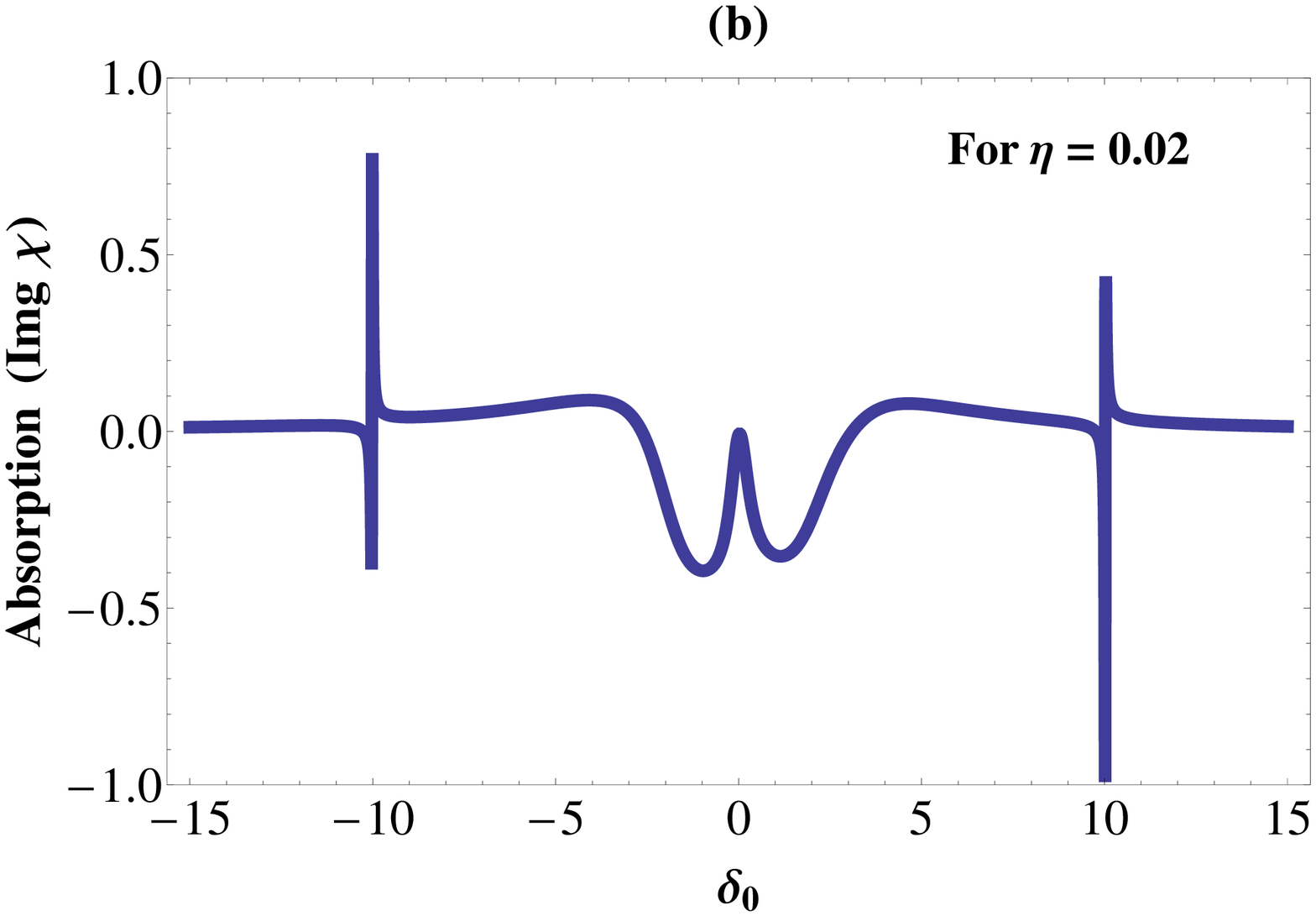} \\
 	\end{tabular}
 	\caption{(Color online) The graph of absorption w.r.t signal pump detuning; For (a)- $\eta$=0, and (b)-$\eta$=0.02; Rest of the Parameters are-$E_{p0}$=3.15, $g_{0}=1.5$, $\omega_{\vec{k}0}$=10, $\kappa_{c0}$=1.35. (All these parameters are dimensionless w.r.t $\Gamma_{2}$)}
 \end{figure}
 
 A familiar absorption curve is visible when $\eta$=0. However, when $\eta$ $\neq$ 0, coupling of the lattice vibrations with the QD brings some new features. Apart from the familiar central absorption curve, two new sharp peaks appear at both side of the absorption spectrum at $\pm \omega_{\vec{k}0}$. These new peaks are due to the lattice vibrations and appear at $\delta_{0}$= $\pm \omega_{\vec{k}0}$. The central peak at $\delta_{0}$ =0 corresponds to Rayleigh resonance. It is interesting to note that for $\eta$=0 (Fig. 4(a)), the Railleigh resonance is symmetric while for $\eta \neq$ 0 (Fig. 4(b)), the Rayleigh resonance develop an asymmetric profile due to lattice vibrations.
 
 Thus by keeping the pump frequency fixed and scanning the signal frequency across the exciton frequency, we can detect the lattice vibrations using this all optical technique.
 
 We now explore the normalized power transmission (T) that can be measured at the output terminal of the device. Fig. 5 displays the graph of $T^{2}$ verses signal-exciton detuning $\Delta_{s}$, for different pump power $E_{p0}$=5 (Fig.5(a)) and $E_{p0}$=7 (Fig.5(b)), where $\Delta_{s}$= $\omega_{s}$-$\omega_{ex}$. The plot clearly displays a sharp absorption dip at $\Delta_{s}$=0 with two asymmetrical dips on both side of $\Delta_{s}$=0. This phenomenon demonstrates an obvious asymmetrical Fano resonance line-shape. As the pump power increases, the absorption dip becomes longer. Fig. 5(c) shows the $T^{2}$ plot verses $\Delta_{s}$ in the absence of exciton-phonon coupling ($\eta$=0). We notice a nearly symmetric curve with no absorption dip at $\Delta_{s}$=0 clearly indicating the fact that the absorption dip is MIA.  
  \begin{figure}[htb]
 	\begin{tabular}{cc}
 		\includegraphics [scale=0.50]{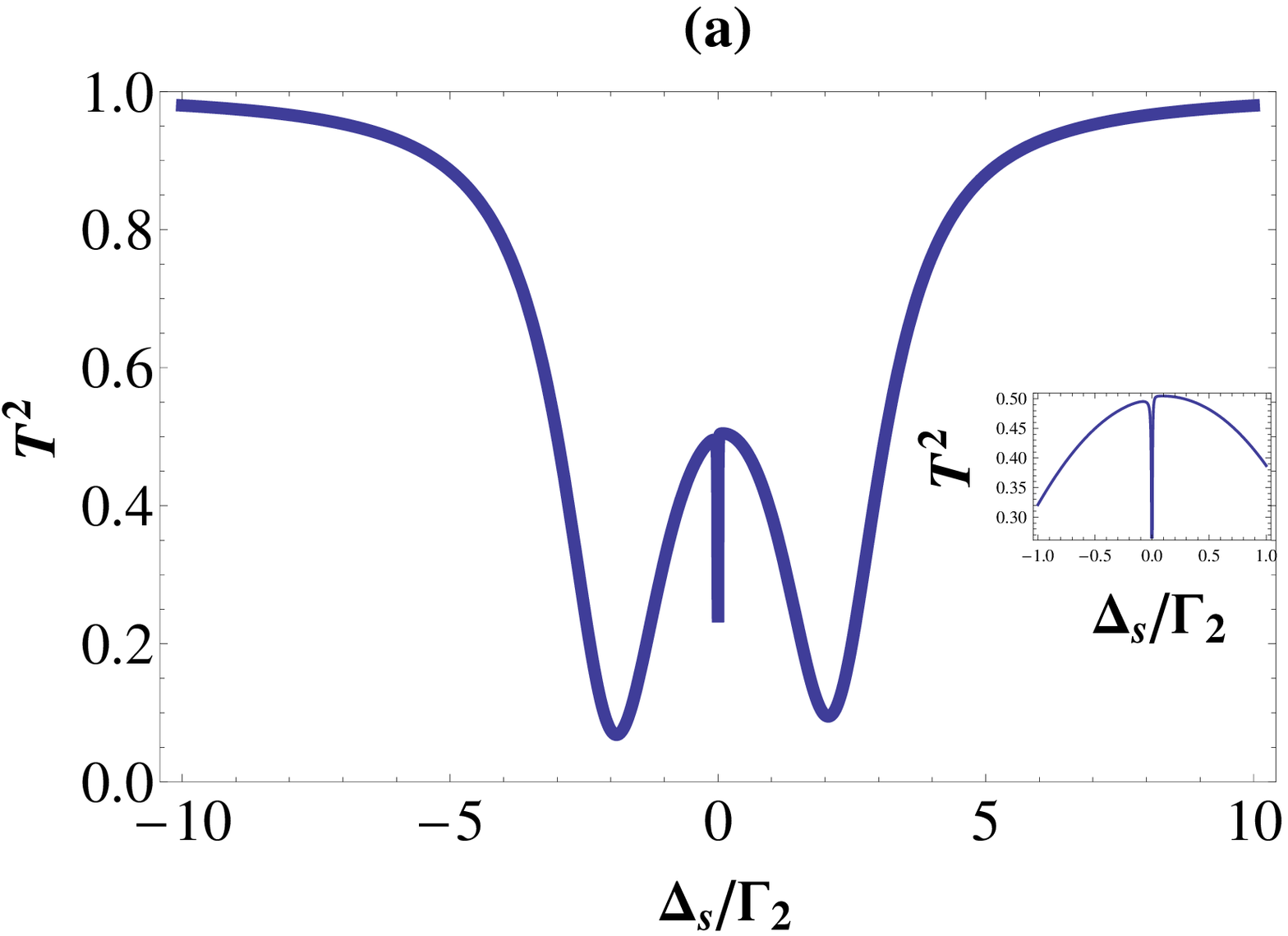} \includegraphics [scale=0.50]{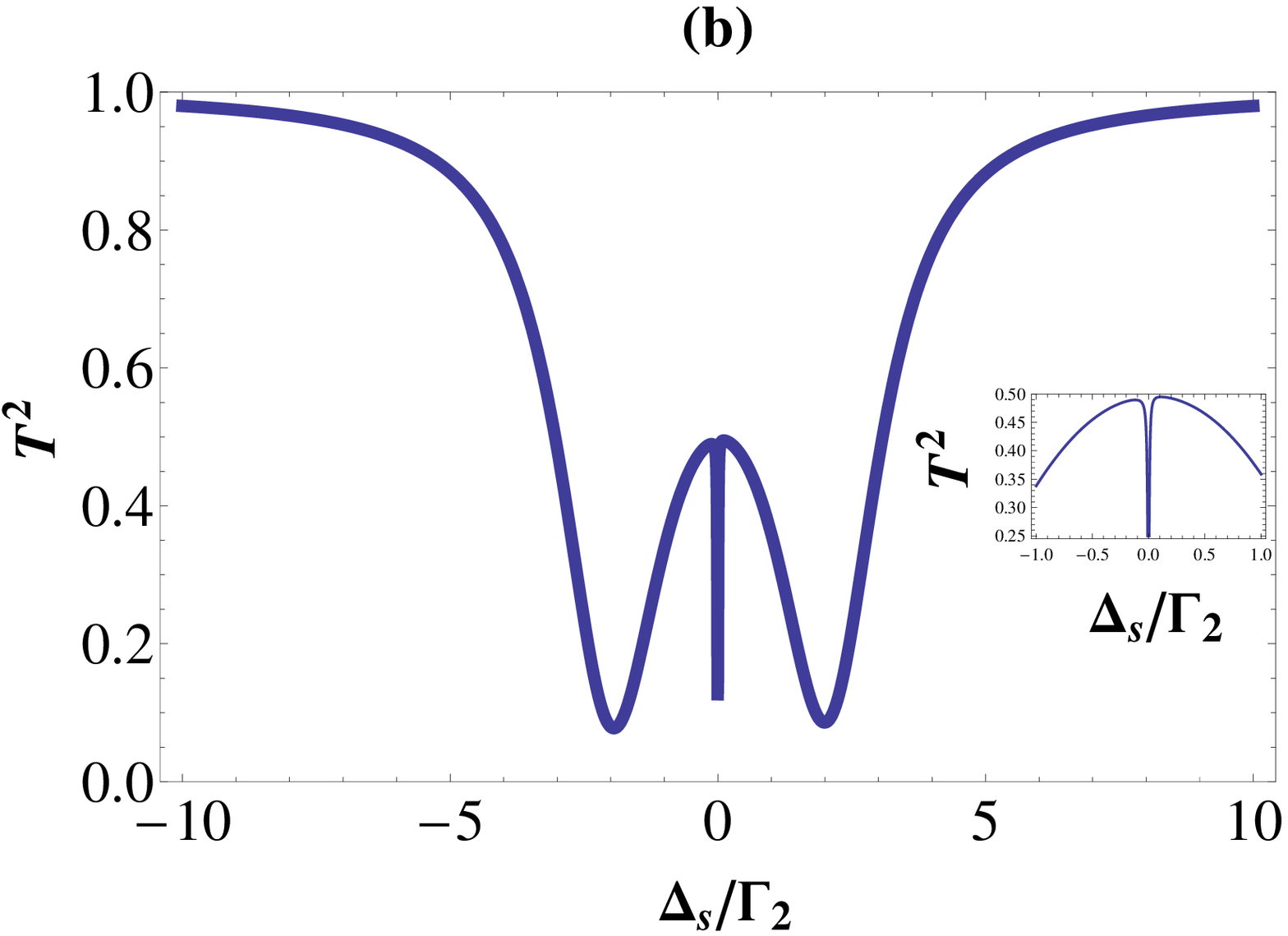}\\ 
 		\includegraphics [scale=0.50] {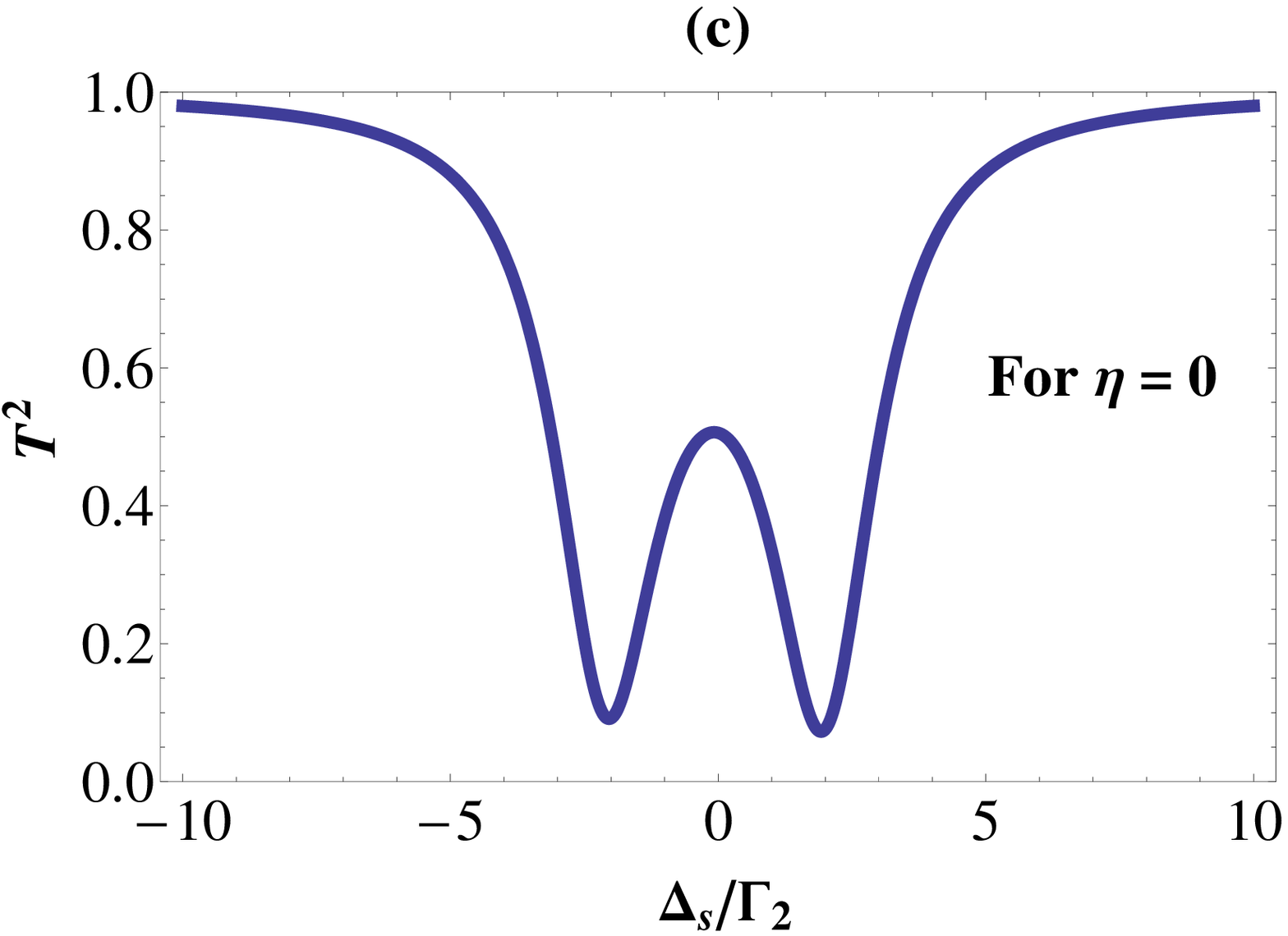}\\
 	\end{tabular}
 	\caption{(Color online)  The graph of Transmission $T^2$ verses signal exciton detuning $\Delta_{s}$ ; Parameters are used for graph are, for (a) $E_{p0}$=5, $\eta$=0.015 (b) $E_{p0}$=7, $\eta$=0.015, (c) $E_{p0}$=5, $\eta$=0; rest all parameters are common- $g_{0}=1.5$, $\omega_{\vec{k}0}$=10, $\kappa_{c0}$=1.35, $\Delta_{c0}$=-10, $\Delta_{p0}$=-10. (All these parameters are dimensionless w.r.t $\Gamma_{2}$)} 
 \end{figure}

 Using Eqn.(19), we plot the real and imaginary part of $a_{out+}$ in fig. 6. The real part Re($a_{out+}$) is the absorption while the imaginary part Im($a_{out+}$) is the dispersive behavior. Fig. 6(a) and 6(b) describes the absorption and dispersive behavior when the exciton is not coupled with the mechanical mode ($\eta=0$). The dispersion curve demonstrates an anomalous behavior around $\Delta_{s}$=0 (near the resonant electronic transition). When the slightly greater radiation frequency interact with refractive index makes it lesser than unity and with increasing the frequency refractive index decreases, this leads to Anomalous dispersion effect. The theory of Anomalous dispersion developed by Lord Rayleigh and demonstrated this phenomenon for a mechanical oscillator \citep{ray-1,ray-2}. Recently in atomic systems electromagnetic interaction due to atomic coherence in degenerate two level systems leads to negative dispersion at the resonant frequency for an atomic transition. In the anomalous dispersion region, the group velocity of light can be negative\citep{akul}.
 
 Fig.6(c) and 6(d) displays the absorption and dispersion when $\eta$=0.015. The results shows that a MIA window appears in $a_{out+}$ around $\Delta_{s}$=0 because the exciton is now coupled to the mechanical mode. At $\Delta_{s}$=0, a very strong absorption is seen together with a highly steep anomalous dispersion around $\Delta_{s}=0$. These results indicate that our system can be used to generate slow light \citep{Ling,akram}. From figure 5 and 6, we also observe that the Fano resonance's transmission contrast can be high (as high as 70 $\%$) which makes it suitable for any Telecom system.  
 
 \begin{figure}[htb]
 	\centering
 	\begin{tabular}{@{}cccc@{}}
 		\includegraphics[width=.50\textwidth]{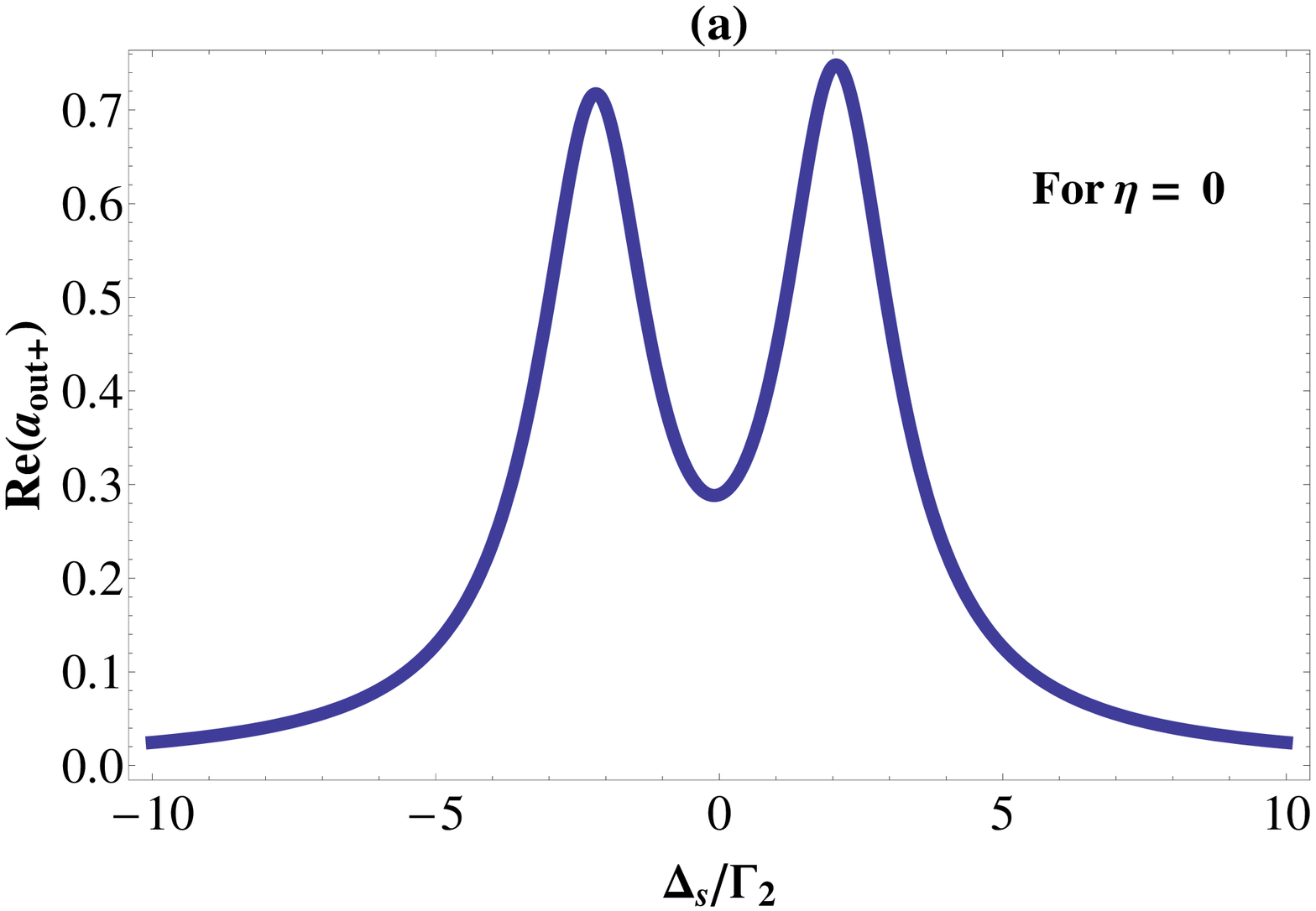} \includegraphics[width=.50\textwidth]{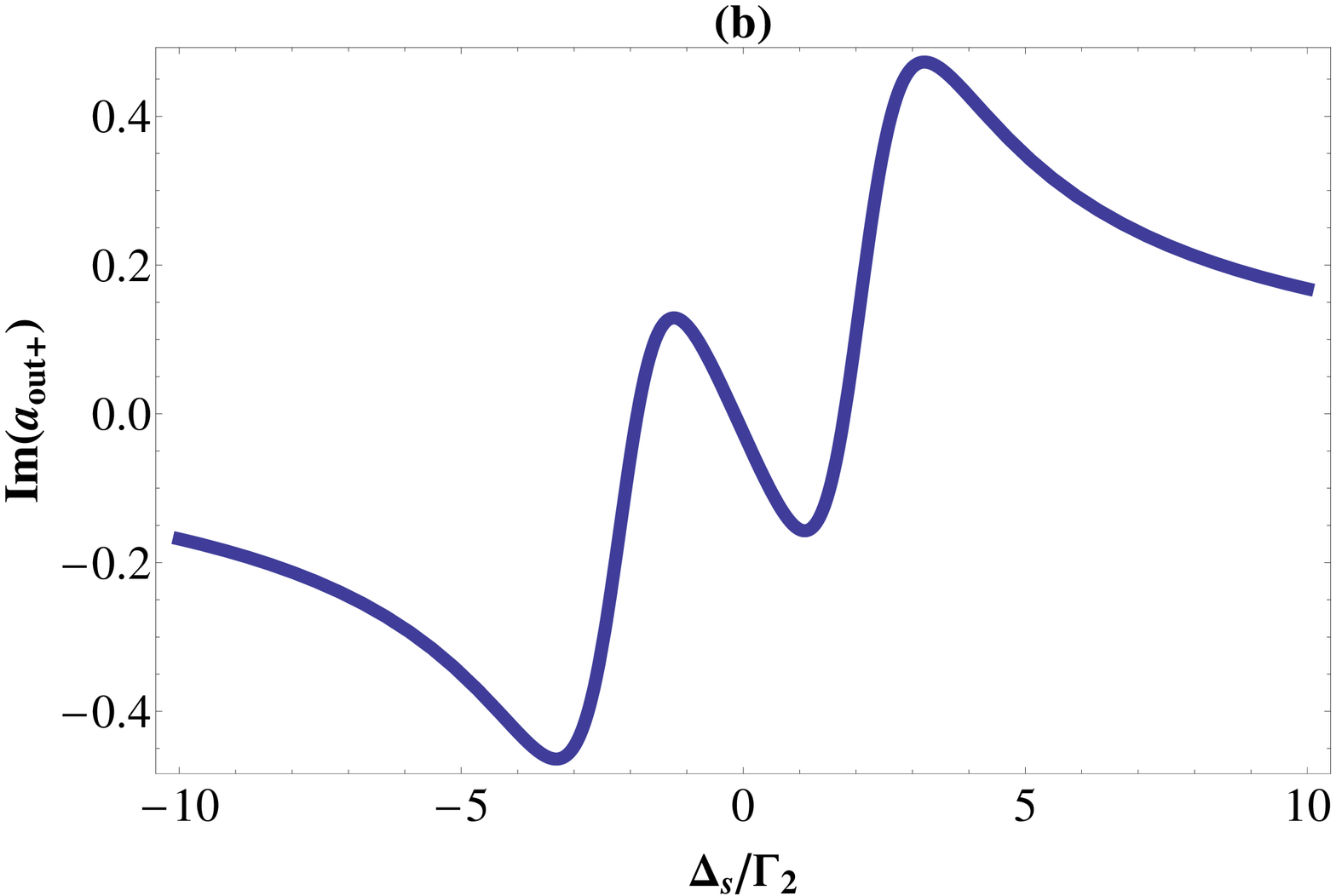}\\
 		\includegraphics[width=.50\textwidth]{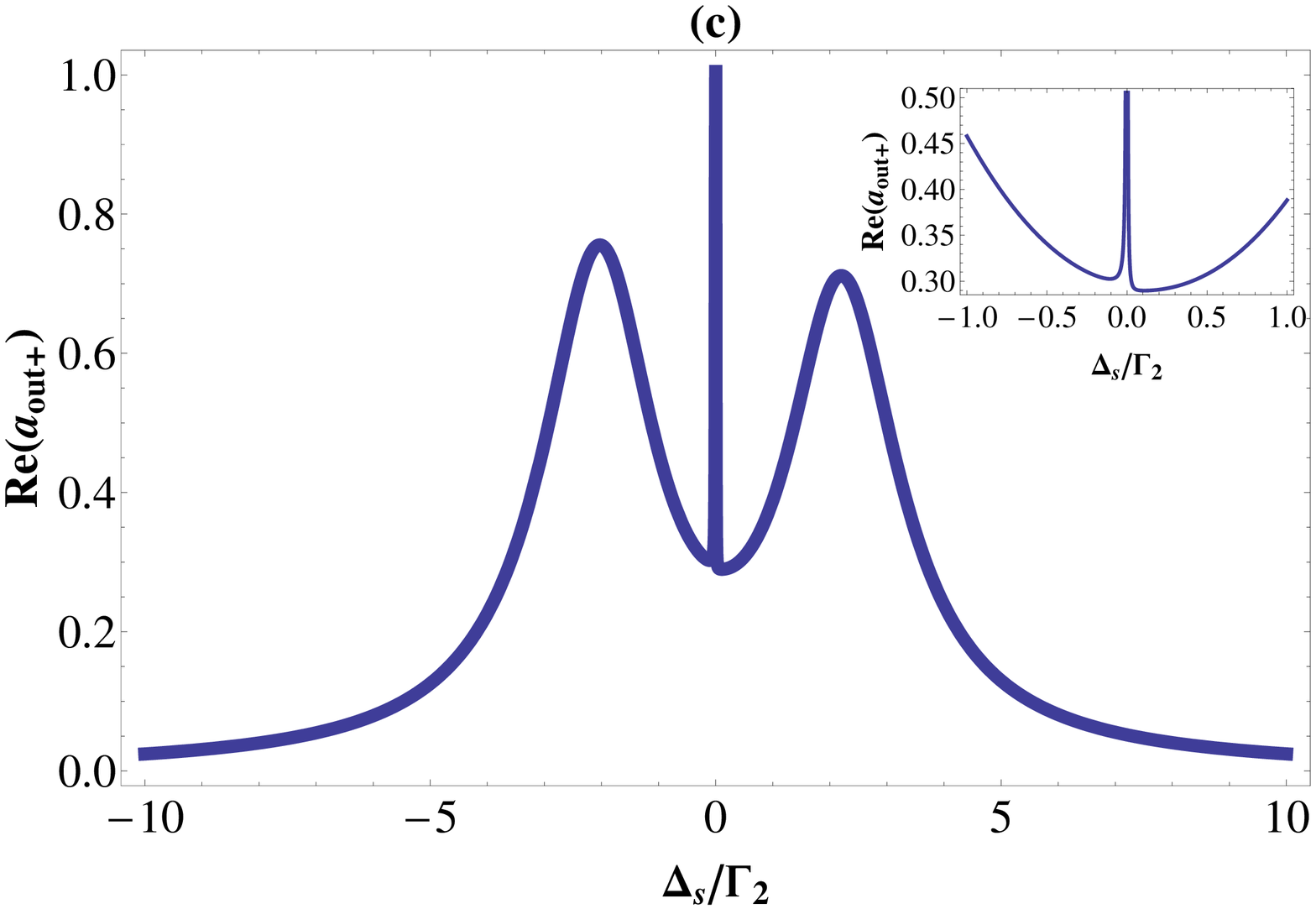} \includegraphics[width=.50\textwidth]{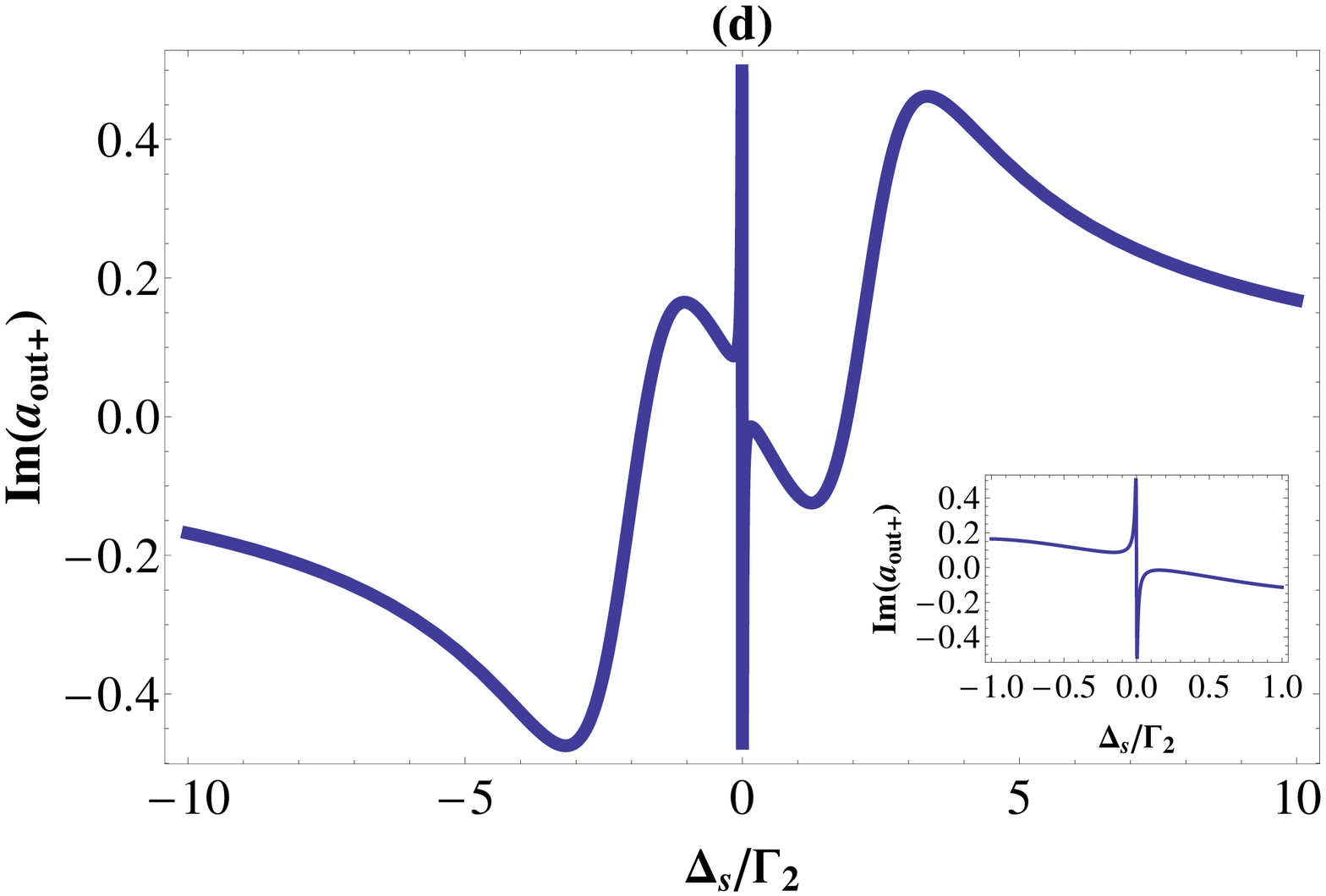}\\
 	\end{tabular}
 	\caption{(Color online): (a)-The graph of Absorption verses signal exciton detuning $\Delta_{s}$ for $\eta$=0; (b)-The graph of dispersion verses signal exciton detuning $\Delta_{s} $ for $\eta$=0 , (c)- for absorption $\eta$=0.015, (d)- for dispersion  $\eta$=0.015; rest of the Parameters are $E_{p0}$=5, $g_{0}=1.5$, $\omega_{\vec{k}0}$=10, $\kappa_{c0}$=1.35, $\Delta_{c0}$=-10, $\Delta_{p0}$=-10. (All these parameters are dimensionless w.r.t $\Gamma_{2}$)}
 \end{figure}
 
 From fig. 7(a) we can note that as the coupling strength between Q.Dot and cavity decreases absorptive behavior increases. And as the value of $g_{0}$ is increase, the fano line shape appears.
 
 \begin{figure}[htb]
 	\centering
 	\begin{tabular}{@{}cccc@{}}
 		\includegraphics[width=.50\textwidth]{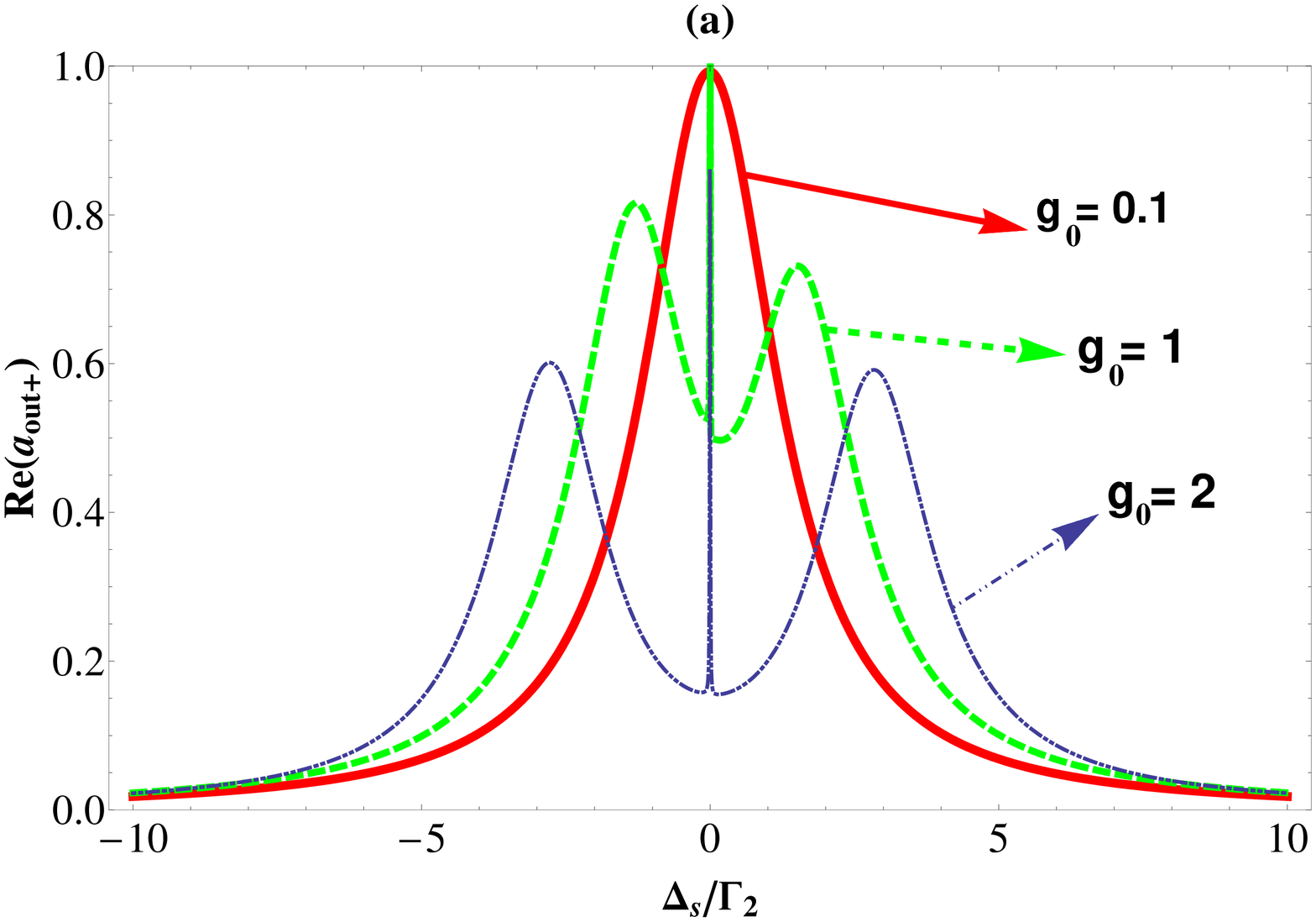} &
 		\includegraphics[width=.50\textwidth]{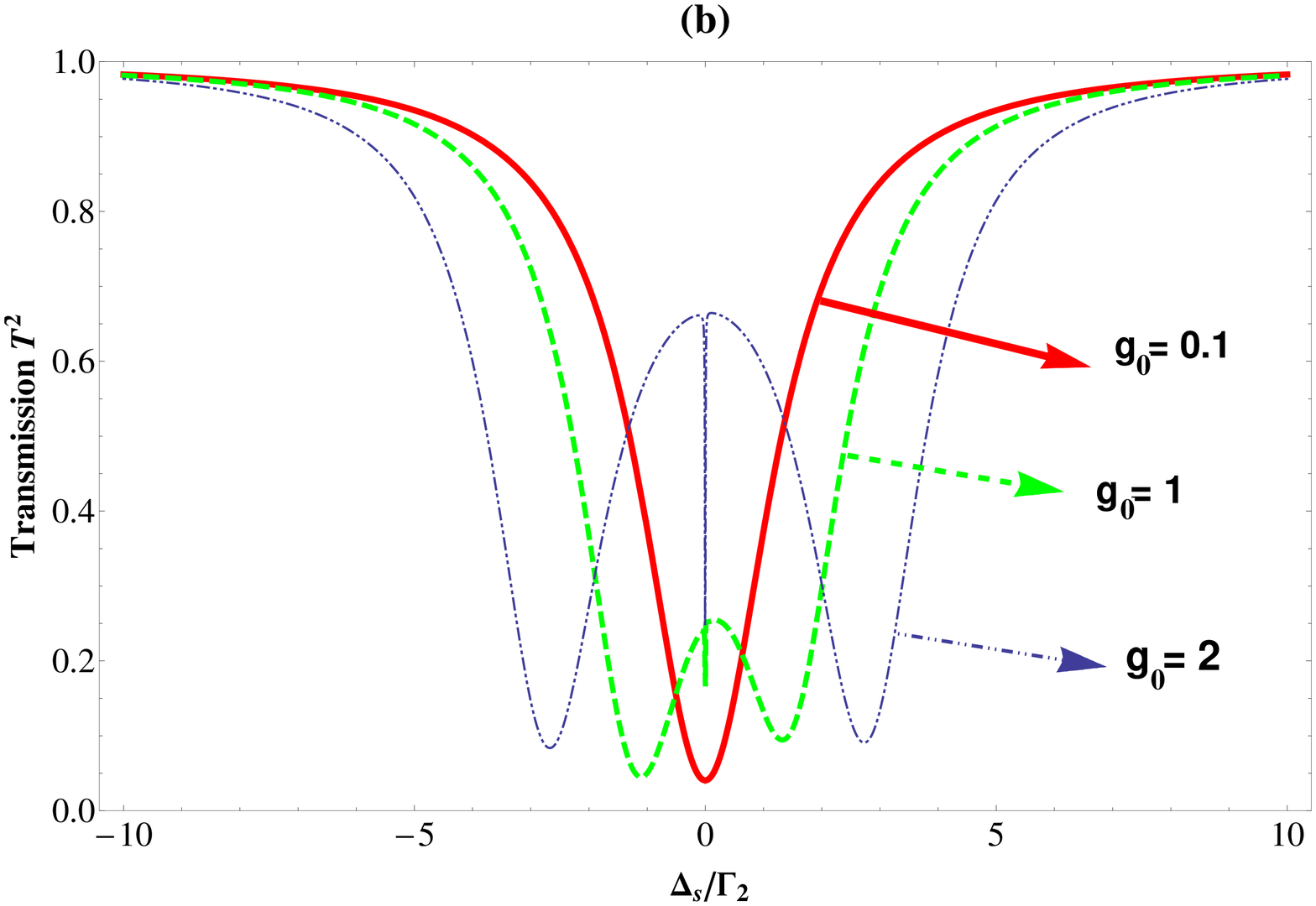}\\
 	\end{tabular}
 	\caption{(Color online): (a)-The graph of absorption $a_{out+}$ w.r.t signal exciton detuning $\Delta_{s}$ for different value of coupling factor. (b)-graph of Transmission $T^2$ w.r.t signal exciton detuning $\Delta_{s}$ for different values of QD-cavity coupling; Parameters for graph are-$E_{p0}$=5 , $\eta$=0.015, $\omega_{\vec{k}0}$= 10, $\kappa_{c0}$=1.35, $\Delta_{c0}$=-10, $\Delta_{p0}$=-10. (All these parameters are dimensionless w.r.t $\Gamma_{2}$)}
 \end{figure}
 
  In this problem we are using two optical field, rather than single field. Fig. 7(b) shows the Transmission spectrum $T^{2}$ as a function of signal-exciton detuning $\Delta_{s}$. We can see that for small coupling between cavity and quantum dot, only a total absorption dip is displayed (Solid line) in the transmission spectrum. and when the coupling is increases gradually the transmission spectrum is splitting into two peaks. We can see that the split width increases while $g_{0}$ increases.
 
 \section{ALL OPTICAL MECHANICAL KERR SWITCH}
 
 The fast development of Internet data traffic poses challenges for optical communication networks. Such networks rely on transferring optical to electronic signals to route information. This imposes limits on the speed and power consumption of transmission. All optical signal routing has the ability to eliminate the need for electrical transduction to overcome these limitations. To implement all optical signals routing, one needs optical switching devices that can function at reduced energy dissipation \citep{arka-b}.

With the rapid development of optics, there is a practical need to understand the nonlinear behavior of the material in order to make efficient use of it when it is wanted and to avoid it when it is unwanted \citep{sem1,has1,lif1,abdel}.
 Many theoretical work has been done so far for the improvement in quality of optical Kerr effect in which optical Kerr effect and control field strength are mutually proportional.\citep{zin}.
 The Kerr effect has attracted a lot of attention in various materials. Application of Kerr effect for telecommunication and future optical information processing, ultra-small and ultrafast active component like all-optical switches is expected \citep{asa}. We use the instantaneous optical Kerr effect for the implementation of ultrafast switches, which is accountable for the self -phase modulation \citep{rob}, nonlinear effects of self-focusing \citep{sod} and modulation instability \citep{tabi}. The quantum dynamics signatures are revealed by this effect \citep{jac}, and adaptive measurement, feedback control are some future applications \citep{jaco}. This paper gives us theoretical understanding of an optomechanical Kerr switch based on lattice vibrations and a quantum dot.
 
 By using equation (15), we get the all optical Kerr effect in the presence and absence of coupling between exciton and phonon as shown in figure(8). We notice from fig.8(a), that the straight(dotted) line representing the absence of Kerr coefficient when the coupling strength ($\eta$=0), and solid line showing the presence of Kerr coefficient when the coupling is present.i.e.,$\eta \ne$0. It means that the exciton-phonon coupling is playing a significant role in this process. When we increase the pump field this enhances the optical Kerr effect (see fig.8(b)). Unlike the red sideband $\Delta_{p0}= \omega_{\vec{k0}}$, here this effect is working on the blue sideband of the pump field $\Delta_{p0}= -\omega_{\vec{k0}}$. The physical source of this outcome is because of coherent population oscillation (CPO), which allows quantum interference between the phonon and two optical field's beat via exciton when the pump-signal detuning $\delta_{0}$ and frequency of phonon are equal $\omega_{\vec{k0}}$. The pump field modulates the scale of Kerr effect.

 In figure(9), the real and imaginary part of $\chi_{eff}^{(3)}$ of equation (15) represents the Kerr coefficient and nonlinear absorption respectively. Clearly, at $\Delta_{s0}$=0 the Kerr coefficient can be significantly enhanced by combining with a positive nonlinear absorption with the detuning $\Delta_{s0}=-\omega_{\vec{k0}}=-10.$

 \begin{figure}[htb]
	\centering
	\begin{tabular}{@{}cccc@{}}
		\includegraphics[width=.50\textwidth]{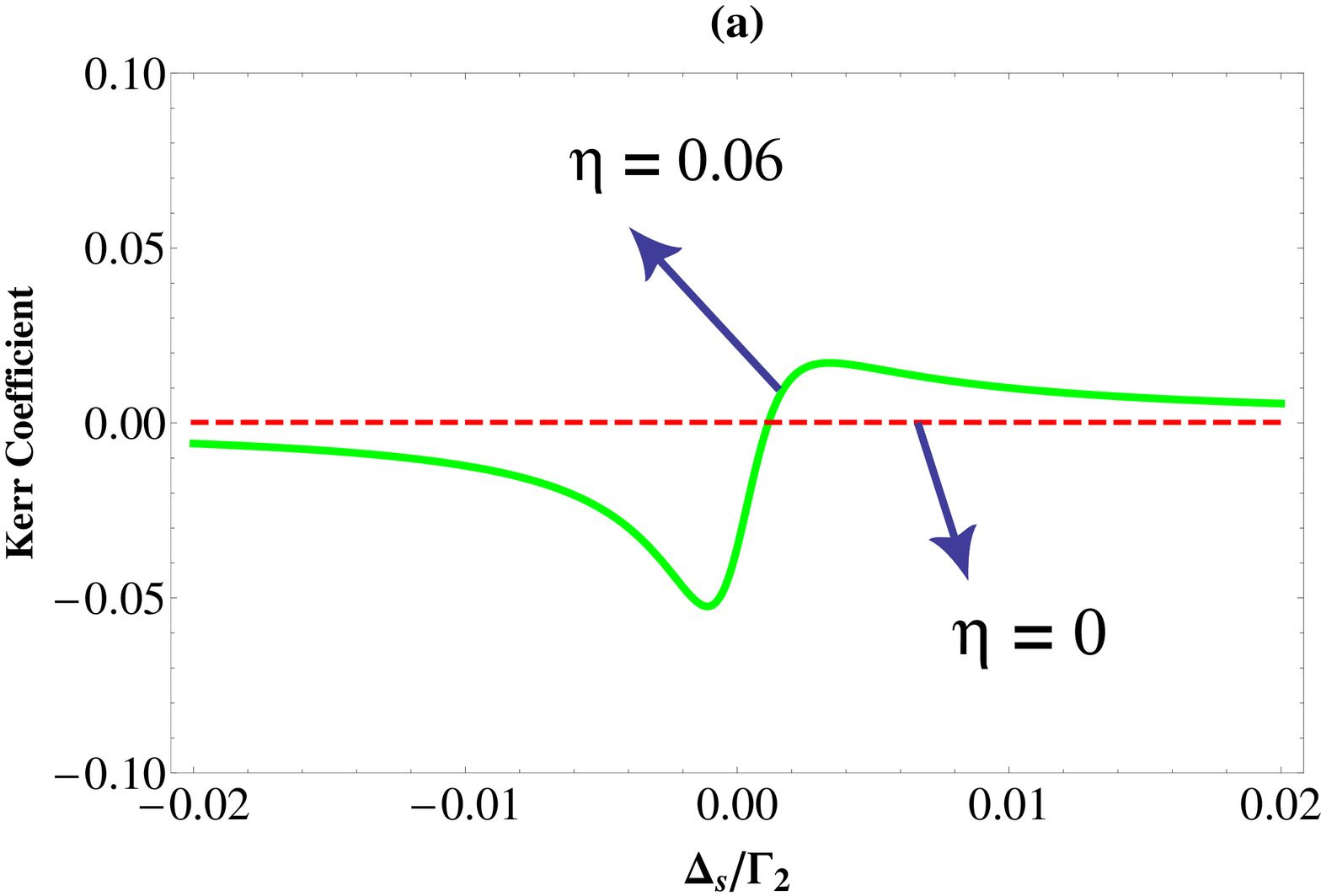}&
		\includegraphics[width=.50\textwidth]{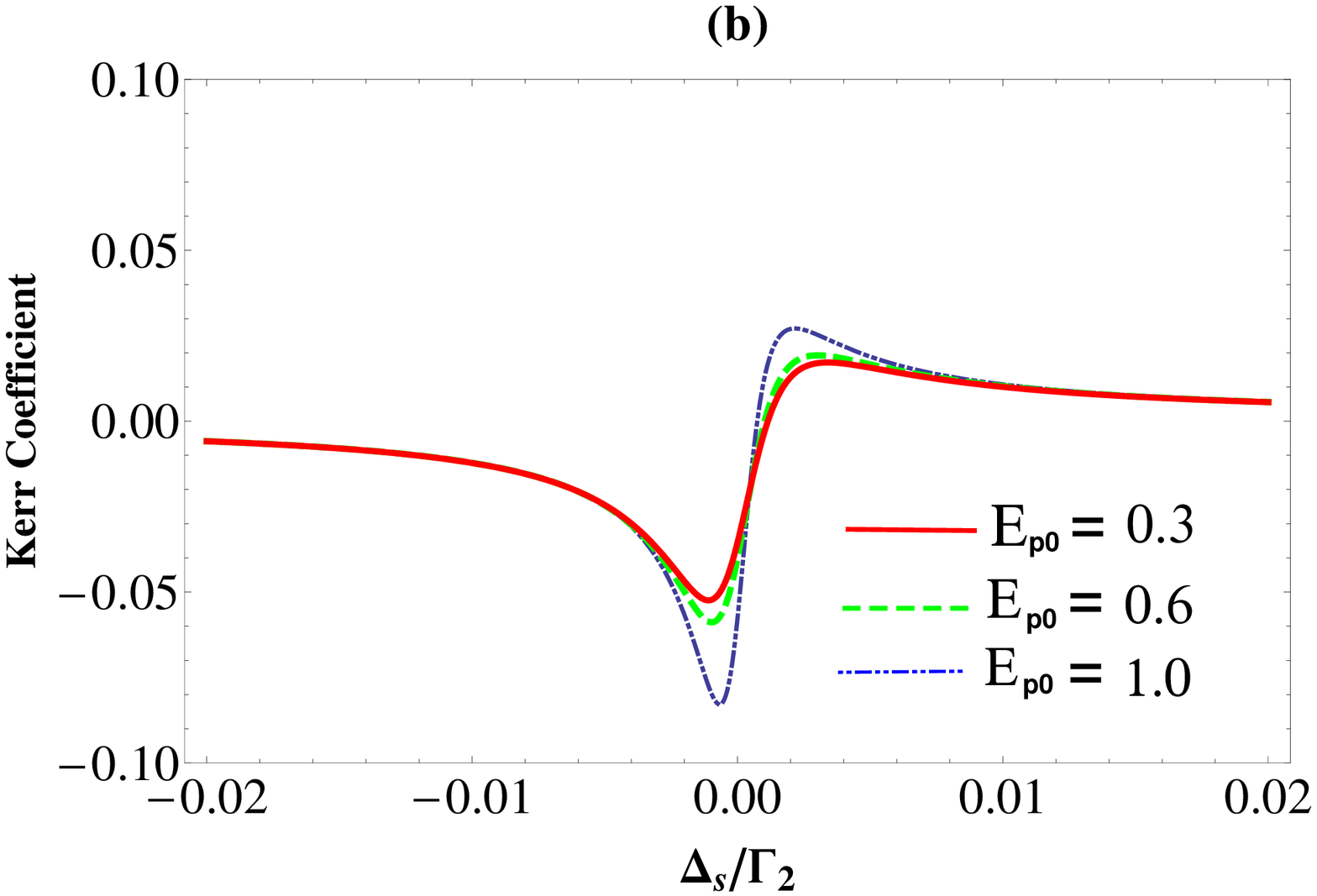} \\
	\end{tabular}
	\caption{(Color online): The graph between optical Kerr coefficient and signal-exciton detuning $\Delta_{s}$; (a)- In presence and absence of coupling between phonon and exciton. (b)- For different value of pump field. Parameters are used for graph are-$E_{p0}$=0.34, $\omega_{\vec{k}0}$=10, $\eta$=0.06, $\kappa_{c0}$=1.35,$\Delta_{c0}$=-10,$\Delta_{p0}$=-10. (All these parameters are dimensionless w.r.t $\Gamma_{2}$)}
\end{figure}
 
 \begin{figure}[htb]
	\centering
	\begin{tabular}{@{}cccc@{}}
		\includegraphics[width=.50\textwidth]{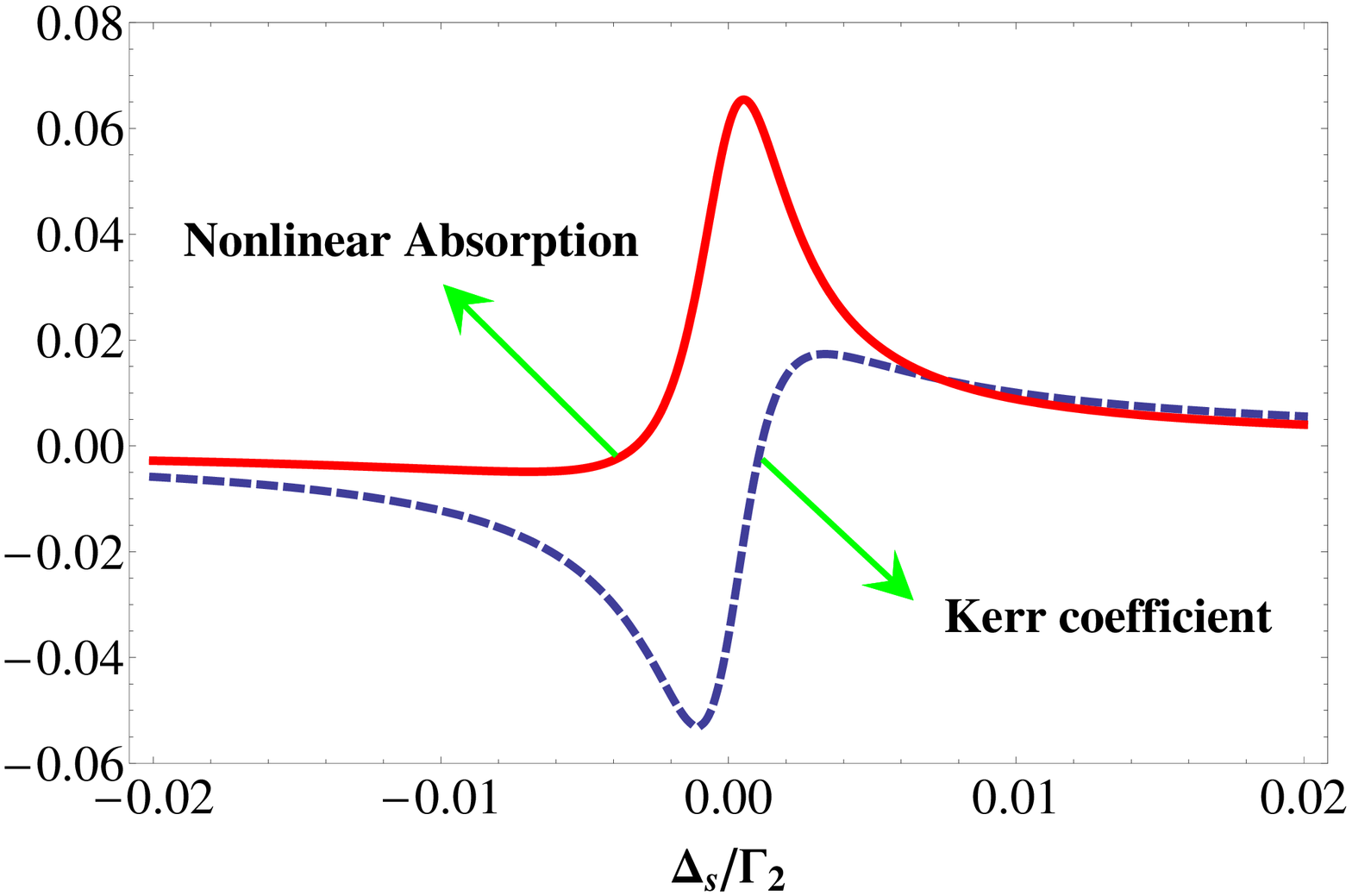}&
		\includegraphics[width=.50\textwidth]{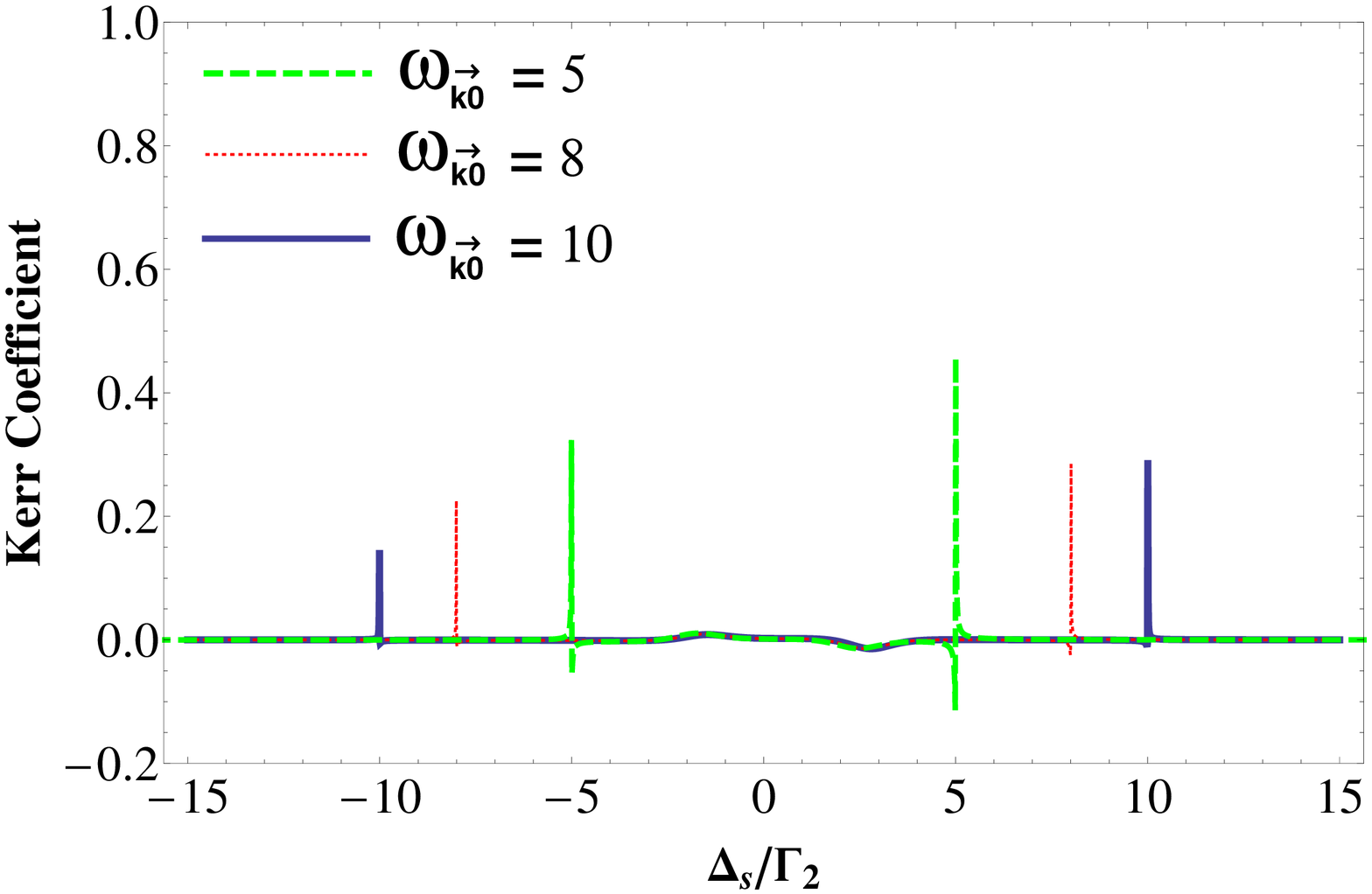} \\
	\end{tabular}
 	\caption{(Color online): (a)- The graph of nonlinear absorption and Kerr coefficient w.r.t signal exciton detuning $\Delta_{s}$. Parameters are used for graph are-$E_{p0}$=0.34, $\omega_{\vec{k}0}$=10, $\eta$=0.06, $\kappa_{c0}$=1.35,$\Delta_{c0}$=-10,$\Delta_{p0}$=-10. (All these parameters are dimensionless w.r.t $\Gamma_{2}$), (b)- The optical Kerr coefficient w.r.t detuning $\Delta_{s}$; Parameters are used for graph are-$E_{p0}$=0.54, $\omega_{\vec{k}0}$=(10,8,5) $\kappa_{c0}$=1.35,$\Delta_{c0}$=0,$\Delta_{p0}$=0. (All these parameters are dimensionless w.r.t $\Gamma_{2}$)}
 \end{figure}

 In the view of above discussion; if we fix the pump-exciton detuning $\Delta_{p0}$=0 and pump cavity detuning $\Delta_{c0}$=0 a different behavior of the Kerr coefficient can be shown in Fig.(10). The two sharply peak at the both side showing the  phonon's vibrational frequency. Like, for the frequency of phonon $\omega_{\vec{k}0}$=10, two sharply peak occurs at $\pm$10 [blue (solid) line]. The red (dotted) line and green (dashed) line are showing two different frequencies of the phonon ($\omega_{\vec{k}0}$)= 8 and 5 respectively, where the sharp peaks analogous to frequency of phonon.

 This implies that, firstly if we fix the control field detuning $\Delta_{p0}$ and $\Delta_{c0}$ and observe the signal frequency in the spectrum across the exciton frequency ($\omega_{ex}$), then we can simply get the phonon's vibrational frequency in nonlinear system precisely.

 \section{CONCLUSION}
 In conclusion we have discussed the optical response of a system comprising of a single quantum dot interacting with the lattice vibrations and the optical mode of a solid state optical micro-cavity. We find that the optical bistability that appears due to the inherent nonlinearity in the system can be tuned by the pump strength and the QD-cavity mode coupling. In particular we show that the bistability appears at lower value of the pump if the QD-cavity mode coupling is high. Further we find that in the signal absorption spectrum, clear signatures of coupling between the QD and the lattice vibrations appears. The absorption spectrum is rendered asymmetric due to the lattice vibrations. An asymmetric Fano resonance line-shape is visible in the Power transmission as a result of exciton-phonon coupling gives rise to anomalous dispersion in the transmitted signal indicating that the system can be used to generate slow light. The Fano resonance's transmission contrast is found to be high making it suitable for Telecom systems. Finally we demonstrate the possibility of using the system as all optomechanical Kerr switch with potential application in fast optical communication networks. 
 
 \section{acknowledgements}
 	\textbf{P.K Jha} and \textbf{Vijay Bhatt} are thankful to \textbf{Department of Science and Technology DST(SERB), Project No. EMR/2017/001980, New Delhi} for the financial support. \textbf{S. A. Barbhuiya} acknowledges \textbf{BITS, Pilani} for the doctorate institute fellowship. \textbf{Aranya B. Bhattacherjee} is grateful to \textbf{BITS Pilani, Hyderabad campus} for the facilities to carry out this research.

\section{APPENDIX-A}
The unknown parameters used in equation(14) are as follows-

\begin{equation}
\tag{A1}
\zeta_{1}(\omega_{\vec{k}})=\frac{\omega^{2}_{\vec{k}0}}{(\omega^{2}_{\vec{k}0} - i \delta_{0}\gamma_{q0} - \delta^{2}_{0})}
\end{equation}

\begin{align}\nonumber
\Phi_{1}&=&(2 - i\delta_{0}) + \frac{2g^{2}_{0}C_{2}\omega_{\vec{k}0}\eta\zeta(\omega_{\vec{k}0})C_{1}}{A_{1}M_{1}} + \frac{2g^{3}_{0}C_{2}D_{1}}{A_{1}M_{1}} +\frac{2 i g_{0}D_{1}\omega_{\vec{k}0}\eta\zeta_{1}(\omega_{\vec{k}0})C_{2}}{N_{1}} + \frac{2 i g^{2}_{0}D_{1}D_{2}}{N_{1}}\\ 
&& +\frac{2g^{2}_{0}C_{1}C_{2}\omega_{\vec{k}0}\eta\zeta_{1}(\omega_{\vec{k}0})}{B_{1}N_{1}} + \frac{2g^{3}_{0}C_{1}D_{2}}{B_{1}N_{1}} - \frac{2 i g_{0}D_{2}\omega_{\vec{k}0}\eta\zeta_{1}(\omega_{\vec{k}0})C_{1}}{M_{1}} - \frac{2 i g^{2}_{0}D_{1}D_{2}}{M_{1}}   \tag{A2} 
\end{align}

\begin{equation}
\tag{A3}
C_{1}=\frac{2g_{0}w_{0}E_{p0}}{\left( i \Delta_{c0}+k_{c0}\right) (\Delta_{p0}-i-2\omega_{\vec{k}0}\eta w_{0}) + 2 i g^{2}_{0} w_{0}}
\end{equation}

\begin{equation}
\tag{A4}
C_{2}=\frac{2g_{0} w_{0} E_{p0}}{\left( -i \Delta_{c0}+k_{c0}\right) (\Delta_{p0}+i-2\omega_{\vec{k}0}\eta w_{0}) - 2 i g^{2}_{0} w_{0}}
\end{equation}

\begin{equation}
\tag{A5}
D_{1}=\frac{-(-w_{0} + i g_{0}C_{1})}{\left(i \Delta_{c0} + k_{c0}\right)}
\end{equation}

\begin{equation}
\tag{A6}
D_{2}=\frac{-(-w_{0} - i g_{0}C_{2})}{\left(-i \Delta_{c0} + k_{c0}\right)}
\end{equation}

\begin{equation}
\tag{A7}
M_{1}=(\Delta_{p0} - i - \delta_{0} - 2\omega_{\vec{k}0}\eta w_{0}) + i \frac{2g^{2}_{0} w_{0}}{\left( i \Delta_{c0} + k_{c0} - i \delta_{0}\right)}
\end{equation}

\begin{equation}
\tag{A8}
N_{1}=(\Delta_{p0} + i + \delta_{0} - 2\omega_{\vec{k}0}\eta w_{0}) -i  \frac{2g^{2}_{0} w_{0}}{\left( -i \Delta_{c0} + k_{c0} - i \delta_{0}\right) }
\end{equation}

\begin{equation}
\tag{A9}
A_{1}=i \Delta_{c0} + k_{c0} -i \delta_{0}
\end{equation}

\begin{equation}
\tag{A10}
B_{1}=-i \Delta_{c0} + k_{c0} - i \delta_{0}
\end{equation}

\section{APPENDIX-B}
The unknown parameters used in equation(16) are as follows-

\begin{equation}
\tag{B1}
\zeta_{2}(\omega_{\vec{k}})=\frac{\omega^{2}_{\vec{k}0}}{(\omega^{2}_{\vec{k}0} + i \delta_{0}\gamma_{q0} - \delta^{2}_{0})}
\end{equation}

\begin{align}\nonumber
\Phi_{2}&=&(2 + i\delta_{0}) + \frac{2g^{2}_{0}C_{2}\omega_{\vec{k}0}\eta\zeta_{2}(\omega_{\vec{k}0})C_{1}}{A_{2}M_{2}} + \frac{2g^{3}_{0}C_{1}D_{2}}{A_{2}M_{2}} -\frac{2 i g_{0}D_{2}\omega_{\vec{k}0}\eta\zeta_{2}(\omega_{\vec{k}0})C_{1}}{N_{2}} - \frac{2 i g^{2}_{0}D_{1}D_{2}}{N_{2}}\\ 
&& +\frac{2g^{2}_{0}C_{1}C_{2}\omega_{\vec{k}0}\eta\zeta_{2}(\omega_{\vec{k}0})}{B_{2}N_{2}} + \frac{2g^{3}_{0}C_{2}D_{1}}{B_{2}N_{2}} + \frac{2 i g_{0}D_{1}\omega_{\vec{k}0}\eta\zeta_{2}(\omega_{\vec{k}0})C_{2}}{M_{2}} + \frac{2 i g^{2}_{0}D_{1}D_{2}}{M_{2}} \tag{B2}
\end{align}

\begin{equation}
\tag{B3}
M_{2}=(\Delta_{p0} + i - \delta_{0} - 2\omega_{\vec{k}0}\eta w_{0}) - i \frac{2g^{2}_{0} w_{0}}{\left( -i \Delta_{c0} + k_{c0} + i \delta_{0}\right)}
\end{equation}

\begin{equation}
\tag{B4}
N_{2}=(\Delta_{p0} - i + \delta_{0} - 2\omega_{\vec{k}0}\eta w_{0}) +i  \frac{2g^{2}_{0} w_{0}}{\left( i \Delta_{c0} + k_{c0} + i \delta_{0}\right) }
\end{equation}

\begin{equation}
\tag{B5}
A_{2}=-i \Delta_{c0} + k_{c0} +i \delta_{0}
\end{equation}

\begin{equation}
\tag{B6}
B_{2}=i \Delta_{c0} + k_{c0} + i \delta_{0}
\end{equation}

\section{APPENDIX-C}

By using standard input-output theory for the coupled system  we can obtained transmitted output field, $a_{out}=a_{in}-\sqrt{k_c}a(t)$, where $a_{out}$ is the output field operator.

\begin{equation}
\tag{C1}
<a_out>= a_{out0}+a_{out+}e^{-i\delta t}+a_{out-}e^{i\delta t}
\end{equation}
\hspace{6.3cm}=$\sqrt{2 k_c}(a_0 + a_{+}e^{-i\delta t}+a_{-}e^{i\delta t}$

\begin{equation}
\tag{C2}
a_{out}= C_{PF}e^{-i\omega_{p}t} + C_{SF}e^{-i\omega_{s}t} - \sqrt{2k_c}a_{-}e^{-i(2 \omega_{p}-\omega_{s})t}
\end{equation}

\begin{equation}
\tag{C3}
C_{PF}= E_{p}-\sqrt{2k_c} a_{0}
\end{equation}

\begin{equation}
\tag{C4}
C_{SF}= E_{s}-\sqrt{2k_c} a_{+}
\end{equation} 

The ratio of output and input field amplitude at the signal frequency can be termed as transmission of the signal field and can be written as,

\begin{equation}
\tag{C5}
T=\left|\frac{C_{SF}}{E_{s}}\right|
\end{equation}
\end{document}